# Cooperative Multiplexing: Toward Higher Spectral Efficiency in Multi-antenna Relay Networks

Yijia (Richard) Fan, Chao Wang, H. Vincent Poor, John S. Thompson

## Abstract

Previous work on cooperative communications has concentrated primarily on the diversity benefits of such techniques. This paper, instead, considers the multiplexing benefits of cooperative communications. First, a new interpretation on the fundamental tradeoff between the transmission rate and outage probability in multi-antenna relay networks is given. It follows that multiplexing gains can be obtained at *any finite* SNR, in full-duplex multi-antenna relay networks. Thus relaying can offer not only stronger link reliability, but also higher spectral efficiency.

Specifically, the decode-and-forward protocol is applied and networks that have one source, one destination, and multiple relays are considered. A receive power gain at the relays, which captures the network large scale fading characteristics, is also considered. It is shown that this power gain can significantly affect the system diversity-multiplexing tradeoff for any finite SNR value. Several relaying protocols are proposed and are shown to offer nearly the same outage probability as if the transmit antennas at the source and the relay(s) were co-located, given certain SNR and receive power gains at the relays. Thus a higher multiplexing gain than that of the direct link can be obtained if the destination has more antennas than the source.

Y. Fan and H. V. Poor are with the Department of Electrical Engineering, Princeton University, Princeton, NJ, 08544, USA. (e-mail: yijiafan@princeton.edu, poor@princeton.edu)

C. Wang and J. S. Thompson are with the Institute for Digital Communications, University of Edinburgh, Edinburgh, EH9 3JL, UK. (e-mail: chao.wang@ed.ac.uk, john.thompson@ed.ac.uk)

This research was supported in part by the U.S. National Science Foundation under Grants ANI-03-38807 and CNS-06-25637.

Much of the analysis in the paper is valid for arbitrary channel fading statistics. These results point to a view of relay networks as a means for providing higher spectral efficiency, rather than only link reliability.

# I. INTRODUCTION

## A. Background

Before describing the contributions of the paper, we first review some basic results from two areas of research on which they are based: Multiple-input Multiple-output (MIMO) communications and cooperative communications.

*1) MIMO communications:* MIMO systems have been extensively studied during the last decade. It is well known that a MIMO system has two advantages over single-input single-output (SISO) systems, namely multiplexing gain and diversity gain. The diversity gain can improve the link reliability, while the multiplexing gain enhances the spectral efficiency. It has been revealed that the tradeoff between diversity and multiplexing gain is a key characteristic of MIMO systems. In [3], the diversity-multiplexing tradeoff (DMT) has been defined assuming that the signal-to-noise ratio (SNR) is *infinitely high*:

*Definition 1 (Infinite-SNR DMT [3]):* Consider a family of Gaussian codes $C_\eta$ operating at SNR $\eta$ and having rates $R_\eta$. Assuming sufficiently long codewords, the multiplexing gain and diversity order are defined as

$$r \triangleq \lim_{\eta \to \infty} \frac{R_\eta}{\log_2 \eta}, \quad \text{and} \quad d \triangleq -\lim_{\eta \to \infty} \frac{\log_2 P_{out}(R_\eta)}{\log_2 \eta}, \tag{1}$$

where $P_{out}(R_\eta)$ is the outage probability corresponding to the transmission rate $R$.

Given any multiplexing gain $r$, the diversity gain $d$ in this scenario describes the rate of decay of the outage probability when the SNR tends to infinity. Graphically, $d$ is approximately equal to the negative slope of the log-log plot of the outage probability versus SNR when the SNR tends to infinity. It is well known that, for Rayleigh fading, the infinite-SNR DMT for an $M_r \times M_t$ point-to-point MIMO system (i.e., a system with $M_t$ transmit antennas and $M_r$ receive

 

antennas) is a piece-wise linear function connecting the points $(k, (M_t - k)(M_r - k))$, $k = 0, \cdots, \min(M_t, M_r)$.

More recently, the analysis of DMT has been extended to the finite SNR scenario [4]. Specifically, the definition is given as follows.

*Definition 2 (Finite-SNR DMT [4]):* The finite-SNR multiplexing gain $r$ and diversity gain $d$ are defined as

$$r = \frac{R}{\log(1 + g\eta)}, \quad \text{and} \quad d(r, \eta) = -\frac{\partial \ln P_{out}(r, \eta)}{\partial \ln \eta} \tag{2}$$

where $g$ denotes an array gain achieved at low SNR, and $P_{out}(r, \eta)$ is the outage probability at rate $R = r \log_2(1 + g\eta)$.

Here the diversity gain quantifies the negative slope of the log-log ratio of outage probability to SNR for *any value* of SNR, instead of the asymptotic slope for infinite SNR. Obviously, the finite SNR DMT approaches the infinite-SNR DMT when $\eta \to +\infty$, but the finite-SNR DMT definition is clearly more general, and also more important in terms of practical implications, as most of the SNR operating points encountered in wireless local area networks (WLANs) and cellular networks are in the range of $-10dB$ to $25dB$.

Generally speaking, link reliability is usually the primary consideration in wireless communications, and thus diversity gain is often of paramount importance. However, once a sufficient link reliability is established, i.e., once the diversity gain increases to a sufficient level, higher spectral efficiency (i.e., multiplexing gain) then rises in importance. In this sense, multiplexing gain is the more important advantage of MIMO systems in terms of enabling higher data rate transmission.

*2) Cooperative communications:* Cooperative communications is a more recent concept that combines the benefits of MIMO systems with relay technologies. In a relay network where the nodes are equipped with either single or multiple antennas, cooperative communications allow the nodes to help each other forward (relay) all messages to the destination, rather than transmitting only their own messages. As the antennas at the transmitters in such network are distributed, the network thus forms a "distributed MIMO" system. A question naturally arises:

How does the performance of a distributed MIMO system compare with that of a point-to-point MIMO system in terms of DMT?

Instead of looking at both multiplexing and diversity behavior simultaneously, most previous work in this area emphasizes primarily the diversity benefits of relay networks (e.g., [1], [2]). The term, *cooperative diversity*, thus has been quoted extensively. The reason for ignoring the multiplexing gain of such networks is primarily because, unlike a point-to-point MIMO link, in a relay network, a multiplexing gain higher than the direct link (i.e., the source to destination link) is difficult to obtain due to the additional receive and transmission time slots the relays require. Several schemes have been proposed to improve the multiplexing gain of half-duplex relay networks [5], [15]–[17]. But still, no multiplexing gain higher than that of the direct link can be obtained in terms of infinite-SNR DMT. In fact, it can be shown [6] that, for a *full-duplex* network consisting of one source, one destination, the infinite-SNR DMT $d(r)$ of the network, at least under *Rayleigh fading*, is upper bounded by

$$d(r) \leq \min\left(d_{\mathcal{S},\mathcal{RD}}, d_{\mathcal{SR},\mathcal{D}}\right), \tag{3}$$

where $d_{\mathcal{S},\mathcal{RD}}$ is the infinite-SNR DMT when the *receive* antennas at the relay(s) and the destination are co-located, and $d_{\mathcal{SR},\mathcal{D}}$ is the infinite-SNR DMT when the *transmit* antennas at the source and the relay(s) are co-located. Clearly, the maximal $r$ of the network is always the same as that of the direct link. [1]

This result is not encouraging. Since the infinite-SNR case is an extreme case of the finite-SNR case, one might conjecture that the same conclusion should apply for the finite-SNR case. However, in this paper, we will show that this conjecture is not true.

### B. Contributions of the paper

In this paper, a new interpretation on the fundamental tradeoff between the transmission rate and outage probability in multi-antenna relay networks is given. It follows that, *under any*

---

[1]Note that the same conclusion is also conjectured in [6] even if the nodes are *clustered*, i.e., when the channel between either the source and the relay or the relay and the destination is an additive white Gaussian noise (AWGN) channel.

*fading statistics*, a higher multiplexing gain than that of the direct link may be obtained by using relaying, for *any finite SNR value*, in multi-antenna relay networks. Thus, cooperative communications can offer not only stronger link reliability, but also higher spectral efficiency as well.

Specifically, we consider a network that has one source, one destination, and multiple relays. We apply the decode-and-forward protocol, where the relays decode, re-encode and forward the message of the source to the destination. The signal received by the relay has average linear power gain $\varphi$ over that received by the destination, where $\varphi$ satisfies $0 < \varphi \leq +\infty$. This power gain $\varphi$ results from the different distances (path losses) between the nodes, and captures the large scale fading characteristics[2] of the network. The gain $\varphi$ is an important factor affecting the multi-node network performance. However, its importance has not been considered fully in previous work. In this paper, we show that the DMT of the network is, in fact, significantly affected by the value of $\varphi$. We summarize the content of the paper as follows.

- For single relay networks, a full-duplex relaying scheme in which the relay uses a fixed set of antennas to re-transmit the message is introduced. Also, an adaptive relaying protocol is proposed based on the principle that the relay transmits only if it decodes the message correctly. This protocol is shown to offer considerable insight into the fundamentals of system outage and DMT performance. Specifically, it is shown that when the SNR is finite, the network DMT performance is mainly determined by the relationship between SNR and network fading characteristics including both $\varphi$ (i.e., large scale fading) and small scale fading coefficients (e.g., Rayleigh fading). The network can offer an outage probability $P_{\mathcal{SR,D}}$ as if the antennas at the source and the relay are co-located, given a sufficiently large $\varphi$. Thus a multiplexing gain higher than that of the direct transmission can be obtained once the destination is equipped with more antennas than the source. It is shown that the higher the SNR becomes, the less likely that $P_{\mathcal{SR,D}}$ can be obtained for a fixed value of $\varphi$. *However*, when the SNR is below a certain threshold, which is determined primarily by $\varphi$, a multiplexing gain that is approximately the same as $d_{\mathcal{SR,D}}^f$ can be obtained with a negligible outage probability, where $d_{\mathcal{SR,D}}^f$ is the finite-SNR DMT corresponding to the outage probability $P_{\mathcal{SR,D}}$.

---

[2] We will also discuss the impact of lognormal shadowing effects in Section VIII.

- For single-relay networks, we examine the *effectiveness* of the proposed encoding and transmission strategies in terms of the probability of providing $P_{\mathcal{SR,D}}$ and $d_{\mathcal{SR,D}}^f$ in the high but finite SNR regime. We define a simple criterion describing effectiveness as a function of the number of transmit and receive antennas at each node.

- We also extend the adaptive protocol and introduce a relay selection scheme in the multiple relay scenario. Their performance is analyzed and compared. It is shown that, unlike the single relay scenario, even the schemes that are defined as *not effective* in the single relay scenario can offer significant performance advantage over the direct link for high but finite SNR values, when combined with an adaptive protocol or relay selection. Moreover, unlike the single relay scenario, the system can almost *always* offer a higher multiplexing gain than direct transmission *regardless of* the value of $\varphi$ and SNR, when the number of relays tends to infinity.

- We further introduce several relaying schemes that can offer higher multiplexing gain than direct transmission when combined with an adaptive protocol, based on either full-duplex or half-duplex configuration. Their performance is discussed and compared. Specifically, it is shown for the first time that the standard half-duplex space-time coding proposed in [1], [2] can also offer a higher multiplexing gain than direct transmission in the finite-SNR regime.

Note that most of the contributions in the paper are valid for *arbitrary fading statistics*. Therefore these results are quite general. The conclusions of this paper offer a different perspective from that obtained via conventional *cooperative diversity* schemes. For example, it is suggested that cooperative (or distributed) coding can not only offer diversity, but can also offer multiplexing gain as well. This suggests a new direction for network coding design, and more generally, for exploiting the benefits of wireless networks, in which higher spectral efficiency, and not just link reliability, can be sought. Thus, the basic concept introduced in the paper can be described as *cooperative multiplexing*.

## C. Related work

Very little related work exists. Some capacity analyses [7], [10], [11] on either scalar or specific fading channels have shown that only under certain receive SNR constraints is it possible to achieve a MIMO rate through full-duplex relaying. However, the outage probability and DMT performance of relay networks in fading environments are not thoroughly exploited or discussed in these papers. The finite-SNR DMT for half-duplex single-antenna relay networks was studied in [8], where only conventional protocols are used with no implications of offering higher multiplexing gain than direct transmission. Other works studying the possibilities of achieving MIMO rates through user cooperation often assume additional bandwidth or special fading conditions among the users (e.g., [13]). To the best of our knowledge, there has been no work to date dealing with the constraints and possibilities of obtaining multiplexing gain for relay networks.

## D. Structure and notation

The paper is organized as follows. In Section II, some mathematical preliminaries are introduced. Section III introduces the system model and assumptions. Section IV focuses on the single relay case where a scheme called fixed relaying is used. Section V extends the results in Section IV to multiple-relay networks. Several other useful relaying schemes are introduced and compared with fixed relaying in Section VI. Discussion and future directions are given in Section VII, and conclusions are drawn in Section VIII.

*Notation*: $A \succeq 0$ means that the matrix $A$ is positive semi-definite; superscripts $T$ and $H$ denote the transpose and complex-conjugate transpose, respectively; $\det X$ denotes the determinant of the matrix $X$; $\mathbf{I}$ denotes the identity matrix; $f(x) \doteq x^a$ means that $\lim_{x \to \infty} \frac{\log f(x)}{\log x} = a$; the set $\bar{\mathcal{X}}$ denotes the complement of the set $\mathcal{X}$; and logarithms are to the base 2.

## II. PRELIMINARIES

The following preliminaries will be frequently used throughout the paper.

                                                                                

*Definition 3 (Exchangeable random variables):* A finite sequence of random variables $X_1, X_2, \cdots, X_N$ is said to be *exchangeable* if, for any finite cardinal number $n$ and any two finite sequences $i_1, ..., i_n$ and $j_1, ..., j_n$, the two sequences $X_{i_1}, X_{i_2}, \cdots, X_{i_n}$ and $X_{j_1}, X_{j_2}, \cdots, X_{j_n}$ have the same probability distribution.

Following this definition, we have the following lemma.

*Lemma 1:* Suppose $X_1, X_2, \cdots, X_N$ is an exchangeable sequence of identically distributed random variables; then,

$$\Pr\left(\max\left(X_1, X_2, \cdots, X_N\right) < x\right) \geq \left[P\left(x\right)\right]^N \tag{4}$$

and

$$\Pr\left(\min\left(X_1, X_2, \cdots, X_N\right) < x\right) \leq 1 - \left[1 - P\left(x\right)\right]^N, \tag{5}$$

where $P\left(x\right)$ denotes the common marginal cumulative distribution function (CDF) of the $X_i$'s. The equalities hold when $X_1, X_2, \cdots, X_N$ are independent and identically distributed (i.i.d.).

*Proof:* Please refer to [22]. ∎

*Lemma 2 (Matrix preliminaries [23]):* The following properties hold for matrices $A$ and $B$: (a) If $A \succeq 0$ and $B \succeq 0$, then $A + B \succeq 0$. (b) If $A = CC^H$, then $A \succeq 0$. (c) $\det(AB) = \det(A)\det(B)$ and $\det(I + AB) = \det(I + BA)$. (d) If $A \succeq 0$ and $B \succeq 0$ and $A - B \succeq 0$, then $\det(A) \geq \det(B)$.

*Lemma 3 (Fischer's inequality [23]):* Suppose that

$$P = \left( \begin{array}{cc} A & B \\ B^H & C \end{array} \right)$$

is a positive definite matrix that is partitioned so that $A$ and $C$ are square and nonempty. Then

$$\det P \leq \left(\det A\right)\left(\det C\right).$$

*Lemma 4:* For any vectors $\mathbf{u}_1, \mathbf{u}_2, \cdots, \mathbf{u}_{m_1}$ and $\mathbf{v}_1, \mathbf{v}_2, \cdots, \mathbf{v}_{m_2}$ belonging to $\mathbb{C}^N$, the follow-

ing inequality holds:

$$\det\left(\mathbf{I}+\sum\mathbf{u}_i\mathbf{u}_i^H\right)\left(\mathbf{I}+\sum\mathbf{v}_i\mathbf{v}_i^H\right)\geqslant\det\left(\mathbf{I}+\sum\mathbf{u}_i\mathbf{u}_i^H+\sum\mathbf{v}_i\mathbf{v}_i^H\right). \qquad (6)$$

*Proof:* See Appendix A. ∎

## III. System Model

### A. General assumptions

It is well recognized that the decode-and-forward protocol is effective only when the relays are close to the source [1], [9]–[12]. Otherwise other relaying modes might be better choices (e.g. compress-and-forward [14]). An underlying reason for this phenomenon is a *power gain* due to shorter paths between the source and the relays compared with the direct link from the source to the destination. With the help of this power gain, reliable communications between the source and relays can be more easily established. However, this path (power) gain has often been ignored in previous work in this area (e.g., [5], [6]). In this paper we will consider the impact of this gain, and will show that it is of considerable importance to the DMT performance for *finite* SNR.

Throughout this paper, we assume a slow, flat, block fading environment, where the channel remains static for a number of transmission time slots. From block to block, the sequences of channel coefficients for the various channels are independent of one another and are individually i.i.d.. We assume that the transmitters do not know the instantaneous channel state information (CSI) on their corresponding forward channels, while CSI is available at the receivers on their receiving channels. We assume that all of the transmit antennas transmit with the same power (i.e., there is no power allocation among the transmit antennas). The white Gaussian noise processes at the receive antennas are assumed to be independent of one another and individually i.i.d. with zero means and unit variances. For any pair of nodes $a$ and $b$, the point-to-point input-output relationship can be expressed as

$$\mathbf{y}=\sqrt{SNR_{a,b}\cdot\varphi_{a,b}}\cdot\mathbf{H}_{a,b}\mathbf{x}+\mathbf{n}_{a,b}$$

where $SNR_{a,b}$ is the power (or SNR, since the noise processes are normalized) per transmit antenna at the transmitter, $\varphi_{a,b}$ denotes the path power gain between nodes $a$ and $b$ due to the effect of the large scale fading in the network such as path loss (this quantity will be described in more detail in Section III.B), $\mathbf{H}_{a,b}$ is the channel transfer matrix between node $a$ and $b$ quantifying the effects of small scale fading, $\mathbf{x}$ is the transmit signal vector with covariance matrix $\mathbf{I}$, and $\mathbf{n}_{a,b}$ is the noise at the receiver. The elements of $\mathbf{H}_{a,b}$ are i.i.d. complex random variables with unit variances, but otherwise can follow *any* distribution. A typical example, of course, is Rayleigh fading, in which the fading coefficients are complex Gaussian random variables with zero means and equal variances $1/2$ for both real and imaginary parts.

## B. Network models

In a single source network, a single source $\mathcal{S}$ is equipped with $K$ antennas, and wishes to communicate with a destination $\mathcal{D}$ that is equipped with $N$ antennas. There are $\mathfrak{R}$ relays in the network, and each relay is equipped with $M_i$ antennas. The relays $\{\mathcal{R}_i\}$ can, for example, be mobile users.

For convenience, we assume unit distance between $\mathcal{S}$ and $\mathcal{D}$, and that the path gain $\varphi_{\mathcal{S},\mathcal{D}}$ is one. Since the relays are assumed to be close to the source, the distance between $\mathcal{R}_i$ and $\mathcal{D}$ is thus approximately the same as that between $\mathcal{S}$ and $\mathcal{D}$, hence $\varphi_{\mathcal{R}_i,\mathcal{D}}$ is also one. Based on the assumption that all transmit antennas transmit with the same power, we define

$$\eta \triangleq SNR_{\mathcal{R}_i,\mathcal{D}} = SNR_{\mathcal{S},\mathcal{D}}. \tag{7}$$

There is also a path gain $\varphi_{\mathcal{S},\mathcal{R}_i}$ between $\mathcal{S}$ and $\mathcal{R}_i$. For notational simplicity, we denote $\varphi_{\mathcal{S},\mathcal{R}_i}$ by $\varphi_i$. The value of $\varphi_i$ can be expressed as a function of $\varphi_{\mathcal{S},\mathcal{D}}$ and the ratio between the distances:

$$\varphi_i \triangleq \left( \frac{d_{\mathcal{S},\mathcal{R}_i}}{d_{\mathcal{S},\mathcal{D}}} \right)^{-\gamma} \varphi_{\mathcal{S},\mathcal{D}} = d_{\mathcal{S},\mathcal{R}_i}{}^{-\gamma}, \tag{8}$$

where $d_{a,b}$ is the distance between node $a$ and $b$, and $\gamma$ is the path loss exponent. In practice $\gamma$ may vary from $2.5$ to $6$ depending on the environment. For example, when $\gamma \approx 4$, $d_{\mathcal{S},\mathcal{R}_i} \approx 0.1$ results in $\varphi_i \approx 40dB$, $d_{\mathcal{S},\mathcal{R}_i} \approx 0.2$ results in $\varphi_i \approx 30dB$, and $d_{\mathcal{S},\mathcal{R}_i} \approx 0.3$ results in $\varphi_i \approx 20dB$ etc. When $d_{\mathcal{S},\mathcal{R}_i} \approx 1$, $\varphi_i \approx 0dB$ and the system model becomes similar as those proposed in previous work (e.g., [1], [2], [5]). Fig. 1 shows an example of the system model.

## IV. Fixed relaying for Single Relay Network

Instead of looking at the multiple relay case directly, we first study the single relay case. The multiple relay case, in many aspects, will be seen as an extension of the single relay case to be discussed in this section. For simplicity, we omit the relay index $i$ which is used for the multiple relay case.

### A. Standard Protocol Revisited

In this section we revisit the standard full-duplex protocol when the block Markov coding principle originally proposed in [9] is used at the relay. We split the source transmission into frames, each of which contains $L$ messages and is transmitted in $L + 1$ time slots. Each transmission time slot $T_{\mathcal{S},\mathcal{R}}$ for the source to relay link is the same as that for the relay to destination link[3] $T_{\mathcal{R},\mathcal{D}}$, which is assumed to have unit length. We assume that the channel is static during each frame transmission and the relay has a total of $M$ antennas for transmission/reception. The relay chooses a set of $M_r$ antennas to listen to the source for the $L$ time slots, and uses a set of $M_t$ antennas out of the $M$ available to re-transmit the message received in the previous time slot to the destination. To ensure that signal quality is optimized at the relay, we assume that the relay uses all of its $M$ antennas to receive the message, i.e., $M_r = M$[4]. Since the subset of $M_t$ antennas is assumed to be fixed for $L + 1$ time slots, we thus term this relaying scheme *fixed relaying*.

In each time slot $i$ ($1 \leq i \leq L$), the source uses a Gaussian codebook to encode a message $\mathfrak{S}_i$ into a codeword vector, and broadcasts it to the relay and destination. The relay decodes the message, and re-encodes the message into another codeword vector, using a different Gaussian codebook that is *independent of* the one used by the source, and transmits it to the destination in time slot $i+1$. The destination, after receiving the message $\mathfrak{S}_i$ in time slots $i$ and $i+1$, decodes it using a maximal likelihood decoder while treating the message $\mathfrak{S}_{i+1}$ as interference. After the $\mathfrak{S}_i$ is decoded, the destination subtracts it from the received signal vector and continues to decode $\mathfrak{S}_{i+1}$ in the next time slot $i+2$. In the following analysis we always assume that $L$ is sufficiently large so that $(L+1)/L \approx 1$.

---

[3]Note that $T_{\mathcal{S},\mathcal{R}}$ can be made unequal to $T_{\mathcal{R},\mathcal{D}}$ if the source transmits the same message when the relay is transmitting [5], [6]. In this paper the source is assumed to transmit a different message when the relay is transmitting, in order to obtain multiplexing gain for any finite SNR. Thus letting $T_{\mathcal{S},\mathcal{R}} \neq T_{\mathcal{R},\mathcal{D}}$ might not be helpful.

[4]As will be discussed later, assuming $M_t = M$ might not necessarily improve the system performance in the finite SNR regime.

If we assume that the source-relay link is perfect, the relay can always correctly decode the message. The capacity for each message transmission is known to be [9], [12]

$$C_{\mathcal{SR},\mathcal{D}} = \log \det \left( \mathbf{I} + \eta \mathbf{H}_{\mathcal{SR},\mathcal{D}} \mathbf{H}_{\mathcal{SR},\mathcal{D}}^H \right) \tag{9}$$

where $\mathbf{H}_{\mathcal{SR},\mathcal{D}} = \left( \begin{array}{cc} \mathbf{H}_{\mathcal{S},\mathcal{D}} & \mathbf{H}_{\mathcal{R},\mathcal{D}} \end{array} \right)$. We omit the derivation here due to its simple extension from the single antenna case in [9], [10], [12]. Clearly, the system in this scenario mimics an $N \times (K + M_t)$ MIMO system. If we assume the source transmits at a rate $R$, it has to satisfy $R \leq C_{\mathcal{SR},\mathcal{D}}$ in order for the message to be correctly decoded at the relay.

Now consider the source-relay link. Its capacity can be written as

$$C_{\mathcal{S},\mathcal{R}} = \log \det \left( \mathbf{I} + \varphi \eta \mathbf{H}_{\mathcal{S},\mathcal{R}} \mathbf{H}_{\mathcal{S},\mathcal{R}}^H \right).$$

In order to correctly decode the message at the relay, the rate $R$ needs to further satisfy $R \leq C_{\mathcal{S},\mathcal{R}}$. Overall, the rate constraint can be written as $R \leq \min\left( C_{\mathcal{S},\mathcal{R}}, C_{\mathcal{SR},\mathcal{D}} \right)$. Otherwise the system will be in outage. In order to ensure that the source-relay channel does not affect the overall system performance, one will need to ensure that $C_{\mathcal{S},\mathcal{R}} > C_{\mathcal{SR},\mathcal{D}}$, i.e.,

$$\det \left( \mathbf{I} + \varphi \eta \mathbf{H}_{\mathcal{S},\mathcal{R}} \mathbf{H}_{\mathcal{S},\mathcal{R}}^H \right) \geq \det \left( \mathbf{I} + \eta \mathbf{H}_{\mathcal{SR},\mathcal{D}} \mathbf{H}_{\mathcal{SR},\mathcal{D}}^H \right). \tag{10}$$

It can be observed that given any $\eta$, whether or not constraint (10) is met is determined only by a relationship among the channel qualities in the network, regardless of the rate $R$. Therefore, it plays a fundamental role in deciding the network performance. Due to the fading statistics, (10) can be satisfied only with a certain probability no matter how large $\varphi$ is. Furthermore, even

 

if (10) holds, the source and the relay transmitters are not aware of it due to the lack of CSI at the transmitters' side. The question now becomes: How does the probability with which (10) holds affect the overall system performance, given the practical scenario in which the CSI is not available at the transmitter?

## B. Adaptive Protocol Offering New Insights

Since the relay knows the source-relay channel condition, it can measure the source-relay channel capacity to decide whether the message can be correctly decoded. In practice, the relay is usually configured to re-transmit the message *only if* it can decode it correctly [1], [2], [5]. Otherwise, direct transmission is assumed. We call such a relay configuration the *adaptive protocol*. Given any rate $R$, the outage probability given by fixed relaying, when the adaptive protocol is applied, can be expressed as

$$P_{out}^{adp} = \Pr\left(C_{\mathcal{S},\mathcal{R}} < R\right) P_{N \times K} + \Pr\left(C_{\mathcal{S},\mathcal{R}} > R\right) P_{N \times (K+M_t)} \tag{11}$$

where $C_{\mathcal{S},\mathcal{R}}$ is the capacity of the source to relay link, and $P_{N \times K}$ is the outage probability of an $N \times K$ MIMO channel (i.e., the source-destination channel) given rate $R$, $P_{N \times (K+M_t)}$ is the outage probability of an $N \times (K + M_t)$ MIMO channel (i.e., the source-plus-relay to destination channel) given rate $R$. Specifically, the finite-SNR DMT under *Rayleigh fading* given by the adaptive protocol might be derived from (11) by a brute-force calculation using the CDF bound for MIMO capacity in [4], which is extremely complicated. To make the analysis simpler, and also to offer clearer insight into the impact of the channel characteristics themselves on the system performance, we build a connection between $P_{out}^{adp}$ and the probability with which (10) holds.

*Theorem 1:* The outage probability $P_{out}^{adp}$ when using the adaptive protocol for the single relay network can be upper bounded as

$$P_{out}^{adp} \le P_c P_{N \times (K+M_t)} + (1 - P_c) P_{N \times K}, \tag{12}$$

where $P_c$ is the probability with which (10) holds.

*Proof:* See Appendix A. ∎

*Remark 1:* The above theorem offers significant insight into the performance of the network for any value of SNR. The outage performance of the network, given any rate $R$, is in fact a linear combination of the contributions of the two MIMO systems, and their weight is decided by the SNR and the network's own characteristics such as path loss and fading, *regardless of* the rate $R$. One can also think that the system achieves the performance of an $N \times (K + M_t)$ MIMO channel with outage probability $1 - P_c$, where the system performance degrades to that of an $N \times K$ MIMO channel. As long as $P_c$ is large, an $N \times (K + M_t)$ MIMO DMT can be approached with only a small outage probability. For any finite value of $\varphi$, the probability $P_c$ decreases to zero as $\eta \to \infty$. In this scenario, the outage probability approaches $P_{N \times K}$. This confirms the results for infinite SNR that the system cannot have a DMT higher than direct transmission. However, for any reasonably large $P_c$, a significant DMT gain over direct transmission will be obtained. Specifically, for any finite value of $\varphi$, once $\eta$ decreases to a certain level, the system performance will approach that of the source-plus-relay to destination channel, which has the DMT of an $N \times (K + M_t)$ MIMO channel.

Following the same method as used in [4] to estimate the finite-SNR DMT of MIMO channels, we further estimate the finite-SNR DMT performance given by the adaptive protocol. Define $w_{N \times (K + M_t)} = P_c P_{N \times (K + M_t)}$ and $w_{N \times K} = (1 - P_c) P_{N \times K}$. By differentiating the right-hand-side (RHS) of (12), the finite-SNR DMT of the single relay network when using the adaptive protocol can be estimated as

$$d_{adp} \approx \frac{w_{N \times (K + M_t)}}{w_{N \times (K + M_t)} + w_{N \times K}} d_{N \times (K + M_t)} + \frac{w_{N \times K}}{w_{N \times (K + M_t)} + w_{N \times K}} d_{N \times K} - \varepsilon d_c, \qquad (13)$$

where $d_{N \times (K + M_t)}$ and $d_{N \times K}$ are the finite-SNR DMTs for $N \times (K + M_t)$ and $N \times K$ MIMO channels, respectively, and where

$$d_c = -\frac{\partial \ln P_c}{\partial \ln \eta}, \varepsilon = \frac{P_{N \times K}}{w_{N \times (K + M_t)} + w_{N \times K}} - 1. \qquad (14)$$

Clearly, we can see that $d_{adp} \approx d_{N \times (K+M_t)}$ when $P_c \approx 1$, and $d_{adp} \approx d_{N \times K}$ when $P_c \approx 0$. Furthermore, it can also be seen that when $P_c$ starts decreasing from 1, the performance typically degrades due to the fact that $P_{N \times K} \gg P_{N \times (K+M_t)}$ and $d_c > 0$. Finally, when $P_c \approx 0$, the contributions of $d_{N \times (K+M_t)}$ and $d_c$ become negligible and $d_{adp} \approx d_{N \times K}$.

### C. The Effectiveness of Relaying

From the above analysis, it is clear that $P_c$ is the key parameter in deciding the system performance. One clearly wants to increase the value of $P_c$ given a practical SNR. This requires a deeper understanding of the constraint (10). In the following we look further into both high SNR and low SNR regimes, while assuming all channel matrices are of *full rank*.

*1) High SNR regime:* At high but finite SNR, we make the following approximation.

$$\log \det \left( \mathbf{I} + \varphi \eta \mathbf{H}_{\mathcal{S},\mathcal{R}} \mathbf{H}_{\mathcal{S},\mathcal{R}}^H \right) \approx \mathcal{M}_{\mathcal{S},\mathcal{R}} \log \varphi \eta + \log \Lambda_{S,R} \tag{15}$$

where $\mathcal{M}_{\mathcal{S},\mathcal{R}} = \min (K, M)$ is considered as the multiplexing gain *upper bound* for the channel $\mathbf{H}_{\mathcal{S},\mathcal{R}}$, and $\Lambda_{S,R} = \prod_{i=1}^{\mathcal{M}_{\mathcal{S},\mathcal{R}}} \lambda_i^{\mathcal{S},\mathcal{R}}$, where $\{\lambda_i^{\mathcal{S},\mathcal{R}}\}$ are the eigenvalues of $\mathbf{H}_{\mathcal{S},\mathcal{R}} \mathbf{H}_{\mathcal{S},\mathcal{R}}^H$. The value of $\Lambda_{S,R}$ can be either $\det \left( \mathbf{H}_{\mathcal{S},\mathcal{R}} \mathbf{H}_{\mathcal{S},\mathcal{R}}^H \right)$ (for $K > N$) or $\det \left( \mathbf{H}_{\mathcal{S},\mathcal{R}}^H \mathbf{H}_{\mathcal{S},\mathcal{R}} \right)$ (for $K < N$). Applying the same approximation to channel $\mathbf{H}_{\mathcal{S}\mathcal{R},\mathcal{D}}$, and defining $\mathcal{M}_{\mathcal{S}\mathcal{R},\mathcal{D}} = \min(K + M_t, N)$, the constraint (10), after some modification, can be expressed as

$$\eta \leq \nu \cdot \varphi^\omega \tag{16}$$

where $\nu$ is a random variable and can be written as

$$\nu = \left( \frac{\Lambda_{\mathcal{S},\mathcal{R}}}{\Lambda_{\mathcal{S}\mathcal{R},\mathcal{D}}} \right)^{\frac{1}{\mathcal{M}_{\mathcal{S}\mathcal{R},\mathcal{D}} - \mathcal{M}_{\mathcal{S},\mathcal{R}}}}, \tag{17}$$

and where $\omega$ can be expressed as

$$\omega = \frac{\mathcal{M}_{\mathcal{S},\mathcal{R}}}{\mathcal{M}_{\mathcal{S}\mathcal{R},\mathcal{D}} - \mathcal{M}_{\mathcal{S},\mathcal{R}}}. \tag{18}$$

Clearly, for any fixed value of $\nu$, increasing either $\varphi$ or $\omega$ can increase the value of $P_c$. However, the value of $\varphi$ is limited in practice (e.g., smaller than $40dB$ if $d_{\mathcal{S},\mathcal{R}} > 0.1$ for $\gamma = 4$) and it is the value of $\omega$ that dominates the range of $\eta$. How to choose the value of $M_t$ to increase the value of $\omega$ while retaining the benefits of relaying is a major issue. To highlight the impact of $\omega$, we make the following definition regarding the effectiveness of relaying.

*Definition 4:* A relaying scheme is said to be *effective* for (finite) high SNR values, if *both* of the following two conditions are satisfied (a) it offers a multiplexing gain higher than $\min(K, N)$ when assuming perfect source-relay channel and $\eta \to +\infty$, and (b) $\omega \geq 1$.

*Remark 2 (The meaning of $\omega \geq 1$):* At higher SNR, the source plus relay to destination link offers an increase in multiplexing gain of at most $\mathcal{M}_{\mathcal{SR},\mathcal{D}} - \mathcal{M}_{\mathcal{S},\mathcal{R}}$ when compared to the source to relay link, given the same diversity gain. In order for the network to have the DMT of the source plus relay to destination link, the path power gain must be the same order ($\mathcal{M}_{\mathcal{SR},\mathcal{D}} - \mathcal{M}_{\mathcal{S},\mathcal{R}}$) to compensate for the lack of multiplexing gain on the source-relay link. Since the order of the path gain from the source to relay link is the same as its multiplexing gain $\mathcal{M}_{\mathcal{S},\mathcal{R}}$, we thus require $\mathcal{M}_{\mathcal{S},\mathcal{R}} \geq \mathcal{M}_{\mathcal{SR},\mathcal{D}} - \mathcal{M}_{\mathcal{S},\mathcal{R}}$.

*Remark 3:* Note that if (b) is not met when a relaying scheme is used, it does *not* mean that it is not able to provide a multiplexing gain than direct transmission for all possible SNR values. It only means that such a scheme might not be effective in providing a better multiplexing gain higher than direct transmission, when the SNR is higher than a value that has the *same order* as $\varphi$ (i.e., $\nu\varphi$). For example, assume $\nu \approx 1$ and $\varphi = 30dB$. If $\omega = 1.2$ for scheme $A$, then (16) is satisfied when $\eta \lesssim 36dB$. If $\omega = 0.8$ for scheme $B$ in the same scenario, then (16) is satisfied when $\eta \lesssim 24dB$ and the range of the SNRs for it to offer the highest possible DMT performance is $12dB$ smaller than that for scheme $A$. In this sense, scheme $B$ is *less effective* for high SNR ranges. However, it can still offer significant performance advantages over direct transmission when $\eta \lesssim 24dB$, as long as condition (a) is satisfied.

Now we study the effectiveness of fixed relaying. Note that it is not meaningful to increase $M_t$ to satisfy $M_t + K > N$, as the maximal multiplexing gain offered by relaying is constrained by the $N$ receive antennas at the destination. Therefore we need to focus only on the value of $\omega$ within the range $M_t \leq N - K$. In this scenario, we obtain the following condition under which fixed relaying is effective:

*Lemma 5:* Suppose $M_t \leq N - K$. Then fixed relaying is effective if

$$M_t \leq 2 \min\left(K, M\right) - K. \tag{19}$$

*Proof:* The proof is straightforward using *Definition 4* and is thus omitted. ∎

The above result offers a way of choosing the value of $M_t$ in order for fixed relaying to be effective at high SNRs. It can be seen that when $2M \leq K$, $\omega$ is always less than 1. Therefore in this scenario relaying is never *effective*.

*Remark 4 (Effective multiplexing gain region):* For $K < N$, the multiplexing gain of relaying over direct transmission for high SNR, assuming that the relay correctly decodes the message, can be approximated as

$$G \approx \frac{\min\left(K + M_t, N\right)}{K}. \tag{20}$$

Now further applying (19), $G$ is approximately upper bounded by

$$G \lesssim \min\left(2, \frac{2M}{K}, \frac{N}{K}\right). \tag{21}$$

We denote this region as the *effective* multiplexing gain region of fixed relaying over *direct transmission* in the high but finite SNR regime for single relay networks. Clearly, it can be observed that relaying can at most *double* the multiplexing gain of direct transmission in all possible antenna sets $\{K, M, N\}$.

*Remark 5 (Variable $\nu$):* In the high SNR regime, $P_c$ can be expressed approximately as $P_c \approx P_\nu = \Pr\left(\nu > \frac{\eta}{\varphi^\omega}\right)$. Note that inequality (16) can be re-written as

$$\frac{\Lambda_{\mathcal{SR},\mathcal{D}}}{\Lambda_{\mathcal{S},\mathcal{R}}} \eta^{\mathcal{M}_{\mathcal{SR},\mathcal{D}} - \mathcal{M}_{\mathcal{S},\mathcal{R}}} \leq \varphi^{\mathcal{M}_{\mathcal{S},\mathcal{R}}}. \tag{22}$$

Statistically, decreasing the value of $M_t$ can significantly decrease the left-hand side (LHS) of (22). The value of $P_\nu$ can be significantly increased in this sense. A study of the probability distribution of $\frac{\Lambda_{\mathcal{SR},\mathcal{D}}}{\Lambda_{\mathcal{S},\mathcal{R}}}$ for specific fading statistics is however beyond the scope of this paper.

*2) Low SNR regime:* The constraint in (10) can be more easily analyzed in the low SNR region. On noting that $x \log e \approx \log(1 + x)$ for small $x$, we have [24]

$$\log \det \left( \mathbf{I} + \varphi \eta \mathbf{H}_{\mathcal{S},\mathcal{R}} \mathbf{H}_{\mathcal{S},\mathcal{R}}^{H} \right) \approx \varphi \eta \sum_{i,j} |h_{ij}|^2 \log e = (M - M_t) K \varphi \eta \cdot \chi_{(M-M_t)K} \log e \quad (23)$$

where

$$\chi_{(M-M_t)K} = \frac{\sum_{i,j} |h_{ij}|^2}{(M - M_t) K}$$

has unit mean. Specifically, for Rayleigh fading, $(M - M_t) K \chi_{(M-M_t)K}$ follows a chi-square distribution with $2K(M - M_t)$ degrees of freedom. Applying the same approximation to each term in (10), the constraint can be re-written as

$$\frac{N(K + M_t)\chi_{N(K+M_t)}}{(M - M_t)\chi_{(M-M_t)K}} \leq \varphi. \quad (24)$$

We can clearly see that (24) is independent of $\eta$, and the probability with which (24) holds is much higher than $P_\nu$ at high SNR for every value of $\varphi$. Specially, for large values of $K$, and $N$, by the strong law of large numbers, $\chi_{N(K+M_t)} \overset{a.s}{=} 1$ and $\chi_{(M-M_t)K} \overset{a.s}{=} 1$. Constraint (24) can thus be re-written as

$$\frac{N(K + M_t)}{(M - M_t)} \leq \varphi, \quad (25)$$

which of course is fulfilled surely once the value of $\varphi$ is large enough.

We note that in a *very* low SNR regime, for any MIMO channel with $M_t$ transmit antennas and $M_r$ receive antennas with an array gain $g = M_t M_r$[5], the diversity gain can be approximated, using $x \log e \approx \log(1 + x)$, as

$$d(r, \eta) \approx -\eta \frac{\partial \Pr \left( M_t M_r \eta \chi_{M_t M_r} < r M_t M_r \eta \right)}{\partial \eta} = 0 \quad (26)$$

---

[5]Note that the array gain $g$ equals $M_r$ if $\eta$ is defined as the total transmit power rather than the transmit power per antenna. For simplicity, in this paper $g$ is always set to the array gain for the direct link for all schemes in all simulations, i.e., $g = KN$, although the array gain of the single relay system is higher than that of direct transmission and is upper bounded by $g = (K + M_t)N$, which is the array gain of an $N \times (K + M_t)$ MIMO system. We ignore the additional power gain when comparing both of them with direct transmission. This is for the sake of simplicity, given that a change in the value of $g$ only shifts the outage probability curve and its impact on the system DMT performance is negligible when the SNR is not very small.

for any value of $r$. Therefore it is not helpful to obtain a multiplexing gain at low SNR by using relaying[6]. We emphasize that the value of the analysis here is to show that the constraint for $\varphi$ becomes *less stringent* as the SNR decreases. Therefore, for every value of $\varphi$, the system is expected to have a much higher probability of performing like an $N \times (K + M_t)$ MIMO system.

## D. Numerical Results

Figs. 2-3 show the probability $P_c$ for different $(K, M, N, M_t)$ under Rayleigh fading. In Fig. 2, $(K, M, N, M_t) = (2, 2, 4, 2)$ and $\omega = 1$, relaying is effective and $P_c$ is large when $\eta$ is of the same order of $\varphi$ (i.e., when $\varphi - \eta \le 10dB$) . The dashed curves in Fig. 2 are the high SNR approximations of $P_c$, i.e., $P_\nu$. It can be seen that the approximations are very close even for moderate SNR values (e.g., $\varphi = 20dB$). In Fig. 3, $(K, M, N, M_t) = (2, 3, 4, 1)$ and $\omega = 2$, relaying is effective, and the value of $P_c$ is large for a much wider range of SNRs (e.g., $\eta \le 30dB$ for $\varphi = 20dB$) and approaches 1 when the SNR is below $20dB$ for $\varphi = 20dB$. This implies that the $3 \times 3$ MIMO DMT can *almost always* be obtained in this scenario.

Fig. 4 shows $P_{out}^{adp}$ for a (2,2,4,2) system, when $r = 2$ and $\varphi$ is either $20dB$ or $30dB$. Note that the slopes of the curves will be significantly lower and steeper as $r$ becomes smaller and a much lower outage probability can be obtained for each SNR[7]. Clearly, when $r = 2$ there is no diversity for the direct transmission. Therefore the curve has no slope and is identically equal to 1. Relaying can form a $4 \times 4$ MIMO system and thus can bring an additional diversity gain of approximately $(4 - 2)^2 = 4$ for high SNRs, if the signals are correctly decoded at the relay. It can be seen that $P_{out}^{adp}$ is the same as $P_{4 \times 4}$ within the SNR range where $P_c \approx 1$ (see also Fig. 2). Clearly, the curve for the upper bounds shown in (12) suddenly start bending from the SNR point where $P_c$ starts to decrease from 1 ($\eta = 4dB$ for $\varphi = 20dB$ and $\eta = 10dB$ for $\varphi = 30dB$), and their values finally approach that of $P_{2 \times 2}$. These observations confirm the conclusions of *Theorem 1*. However, the exact $P_{out}^{adp}$ is much less sensitive to the value of $P_c$. It can be seen that the curve for the exact value of $P_{out}^{adp}$ bends much more slowly as $P_c$ decreases from 1 and finally becomes a horizontal line. This confirms the conclusion that once $P_c$ approaches zero, the system DMT will be the same as that of direct transmission, which has no diversity gain for rate $r = 2$ in this example.

---

[6]Note that this does not mean that relaying is not helpful in the low SNR regime. The advantage of using relaying in this scenario appears as a power gain, rather than as a multiplexing gain.

[7]Note that, however, most practical wireless systems do not require an error probability of lower than $10^{-3}$. For example, systems such as wireless local area networks (WLANs) have moderate target packet error rates (PERs) around $10^{-2} - 10^{-1}$ [25].

## V. Fixed Relaying for Multiple Relay Networks

The analyses in the single relay scenario can be readily extended to the multiple relay scenario. However, unlike the single relay scenario, more degrees of freedom can be obtained from the multiple source to relay links. This implies that the impact of individual link quality (such as the parameter $\omega$ proposed for the single relay network) is less important. Instead, how to exploit the additional dimension and degrees of freedom offered by multiple relays becomes a major concern.

### A. Multi-casting

For large $L$, the outage probability of the network, on assuming all $\Re$ relays are used to forward the message can be approximated as

$$
\begin{aligned}
P_{mul} &\approx \mathrm{Pr}\left(\min\left\{C_{\mathcal{S},\mathcal{R}_i}\right\} < R\right) + \left(1 - \mathrm{Pr}\left(\min\left\{C_{\mathcal{S},\mathcal{R}_i}\right\} < R\right)\right) P_{N\times\left(K+\sum\limits_{i=1}^{\Re} M_i\right)} \\
&= 1 - \prod_{i=1}^{\Re} \mathrm{Pr}\left(C_{\mathcal{S},\mathcal{R}_i} > R\right) + P_{N\times\left(K+\sum\limits_{i=1}^{\Re} M_i\right)} \prod_{i=1}^{\Re} \mathrm{Pr}\left(C_{\mathcal{S},\mathcal{R}_i} > R\right) \\
&\stackrel{\Re\to\infty}{=} 1
\end{aligned}
\tag{27}
$$

This result is not surprising, as it indicates that cooperative multiplexing gain is difficult to obtain due to the requirement of perfect decoding *at all relays*. Therefore, more advanced protocols are required in order to exploit the advantages of multiple relays.

## B. Optimal Relay Selection

Instead of using all the relays, one may want to use the relay that has the highest source to relay channel capacity to decode and forward the message[8]. This scheme might offer less multiplexing gain, while having a much higher probability of obtaining it. Specifically, on assuming all the relays use the same number of antennas $M_t$ to re-transmit the message if they are chosen, the outage probability can be expressed as

$$
\begin{aligned}
P_{mul}^{ors} &\approx \Pr\left(\max\left\{C_{\mathcal{S},\mathcal{R}_i}\right\} < R\right) + \left(1 - \Pr\left(\max\left\{C_{\mathcal{S},\mathcal{R}_i}\right\} < R\right)\right) P_{N\times(K+M_t)} \quad (28)\\
&= \prod_{i=1}^{R}\Pr\left(C_{\mathcal{S},\mathcal{R}_i} < R\right) + \left(1 - \prod_{i=1}^{R}\Pr\left(C_{\mathcal{S},\mathcal{R}_i} < R\right)\right) P_{N\times(K+M_t)}\\
&\stackrel{\mathfrak{R}\to\infty}{=} P_{N\times(K+M_t)}. \quad (29)
\end{aligned}
$$

Therefore, the outage probability or finite-SNR DMT when using the optimal relay selection scheme is *almost always* similar to that of an $N\times(K+M_t)$ MIMO channel, when the number of relays tends to infinity. More specifically, to offer an insight into the impact of the channel fading characteristics, we have the follow theorem.

*Corollary 1:* The outage probability for fixed relaying with optimal relay selection, on assuming each relay uses $M_t$ antennas to transmit if chosen, is upper bounded by

$$
P_{mul}^{ors} \leq (1 - P_c^{ors}) + P_c^{ors} P_{N\times(K+M_t)}, \quad (30)
$$

where $P_c^{ors} = 1 - \Pr\left(C_{\mathcal{S},\mathcal{R}_i} < C_{\mathcal{S}\mathcal{R}_i,\mathcal{D}}, i = 1, \cdots, \mathfrak{R}\right)$ and $C_{\mathcal{S}\mathcal{R}_i,\mathcal{D}}$ denotes the capacity of the $\mathcal{S}$ plus $\mathcal{R}_i$ to $\mathcal{D}$ channel obtained when the transmit antennas at $\mathcal{S}$ and $\mathcal{R}_i$ are co-located.

*Proof:* The proof is an extension of that of *Theorem 1* and is thus omitted. ∎

Clearly, $P_c^{ors}$ becomes larger as $\mathfrak{R}$ increases, as the scheme requires only one source to relay channel to be sufficiently good.

*Remark 6:* This scheme is especially attractive for those relaying schemes that are *not effective* for single relay networks. Note that the value of $P_c$ can be significantly enlarged to $P_c^{ors}$ in the multiple relay scenario. This means that the SNR range for which the probability of achieving outage probability $P_{N \times (K+M_t)}$ is high is significantly enlarged after performing optimal relay selection. Therefore, even those schemes that have low values of $\omega$ (or $P_c$) can be expected to offer a significant performance advantage in the high but finite SNR regime in the multiple antenna scenario. Thus, the notion of effectiveness defined for the single relay scenario (i.e., *Definition 4*) is no longer that important here.

---

[8]A practical method for choosing the best relay can be found in [21].

### C. Adaptive Protocol

Similarly to the single relay case, an easier way to improve performance is to let the relays re-transmit only if they can decode the message correctly. For the sake of notational simplicity, we assume that $\varphi_i \approx \varphi$, $M_i = M$, $M_{t_i} = M_t$ and $\Pr\left(C_{\mathcal{S},\mathcal{R}_i} < R\right) = P_{\mathcal{S},\mathcal{R}}$ for all $i$. Extensions to more general cases are straightforward. The outage probability in this scenario can be expressed as

$$P_{mul}^{adp} \approx \sum_{i=0}^{\mathfrak{R}} P_i P_{N \times (K+iM_t)}, \tag{31}$$

where

$$P_i = \begin{pmatrix} \mathfrak{R} \\ i \end{pmatrix} \left(P_{\mathcal{S},\mathcal{R}}\right)^{\mathfrak{R}-i} \left(1 - P_{\mathcal{S},\mathcal{R}}\right)^i.$$

Clearly, for each $P_{N \times (K+iM_t)}$ (i.e., an $N \times (K+iM_t)$ MIMO DMT), there is a probability $P_i$ of meeting it, and $\sum_{i=0}^{\mathfrak{R}} P_i = 1$. The overall performance is the combination of the contributions from each effective MIMO system.

One can also formulate the relationship between $P_{mul}^{adp}$ and the network fading characteristics such as the results in *Theorem 1*. However, the analysis is much more complicated here due to the possibility of obtaining different MIMO DMTs. Specifically, for any $i$ relays forming a set $\mathcal{O}_i$, we have the following sufficient condition to ensure that the outage probability is dominated by the destination decoding error (see the analysis in Section IV.A):

$$C_{\mathcal{S},\mathcal{R}_j} \geq C_{\mathcal{S}\mathcal{O}_i,D}, \forall j \in \mathcal{O}_i \tag{32}$$

where

$$C_{\mathcal{SO}_i, \mathcal{D}} = \log \det \left( \mathbf{I} + \eta \sum_{j \in \mathcal{O}_i} \mathbf{H}_{\mathcal{R}_j, \mathcal{D}} \mathbf{H}_{\mathcal{R}_j, \mathcal{D}}^H + \eta \mathbf{H}_{\mathcal{S}, \mathcal{D}} \mathbf{H}_{\mathcal{S}, \mathcal{D}}^H \right).$$

Note that similar to the single relay case, this expression can be significantly simplified in the high SNR regime. Here, we offer an insight into the relationship between the network fading statistics themselves and the probability of obtaining each MIMO DMT that is larger than that of direct transmission (i.e., the probability of obtaining $P_{N \times (K + iM_t)}$ for $i = 1, \ldots, \mathfrak{R}$).

*Theorem 2:* Denote by $P_{c_i}^{\emptyset}$ the probability with which condition (32) is *not* met for any set $\mathcal{O}_i \subseteq \{\mathcal{R}_k\}, k = 1, \ldots, \mathfrak{R}$, i.e., $P_{c_i}^{\emptyset} = \Pr \left( C_{\mathcal{S}, \mathcal{R}_j} < C_{\mathcal{SO}_i, \mathcal{D}}, \exists j \in \mathcal{O}_i, \forall \mathcal{O}_i \subseteq \{\mathcal{R}_k\} \right)$. Then each $P_i$ $(i = 1, \ldots, \mathfrak{R})$ in (31) can be upper bounded by

$$P_i \leq P_{c_i} \tag{33}$$

where $P_{c_{\mathfrak{R}}} = 1 - P_{c_{\mathfrak{R}}}^{\emptyset}$, and

$$P_{c_i} = \begin{cases} 1 - \dfrac{P_{c_i}^{\emptyset}}{P_{c_{i+1}}^{\emptyset}} & \text{for } P_{c_{i+1}}^{\emptyset} \neq 0, \\ 0 & \text{for } P_{c_{i+1}}^{\emptyset} = 0, \end{cases} \quad i = 1, \cdots, \mathfrak{R} - 1. \tag{34}$$

*Proof:* See Appendix C. ∎

*Remark 7:* Compared with (31), in (33) each $P_i$ (except for $i = 0$) is replaced by $P_{c_i}$, i.e., a probability that is related only to the SNR and the channels' own characteristics, *regardless of* the rate $R$. Although the bound (33) might not be tight, it reflects the *weight* of the impact of each MIMO DMT on the system performance. For sufficiently small SNR values, $P_{c_{\mathfrak{R}}}^{\emptyset} \approx P_{c_{\mathfrak{R}-1}}^{\emptyset} \approx \cdots \approx P_{c_1}^{\emptyset} \approx 0$. Thus $P_{c_{\mathfrak{R}}} \approx 1$, and all other $P_{c_i}$'s are zero. The system in this scenario offers an $N \times (K + \mathfrak{R}M_t)$ MIMO DMT. After the SNR increases to a certain point, the value of $P_{c_{\mathfrak{R}}}^{\emptyset}$ starts increasing rapidly towards 1 and the weight of the $N \times (K + \mathfrak{R}M_t)$ MIMO DMT decreases to zero, while $P_{c_{\mathfrak{R}-1}}^{\emptyset}$ is still approximately equal to zero (note that $P_{c_{i-1}}^{\emptyset} < P_{c_i}^{\emptyset}$). In this scenario $P_{c_{\mathfrak{R}-1}} \approx 1$ and the performance is determined mainly by an $N \times (K + (\mathfrak{R} - 1)M_t)$ MIMO DMT. Similar observations can be made for the other values $\{P_{c_i}\}$ as SNR continues increasing, and the system finally offers only the performance of direct transmission (i.e., an $N \times K$ MIMO DMT) for sufficiently large SNR. Overall, as the SNR increases, the system performance will degrade gradually from that of an $N \times (K + \Re M_t)$ MIMO DMT to an $N \times K$ MIMO DMT.

## D. Performance Comparison

*Corollary 2:* On assuming $\varphi_i \approx \varphi$, $M_i = M$, and $M_{t_i} = M_t$, for all $i$, we have

$$P_{mul}^{adp} \lesssim P_{mul}^{ors}$$

.

*Proof:* The result follows since

$$P_{mul}^{adp} \lesssim P_0 P_{N \times K} + \sum_{i=1}^{\Re} P_i P_{N \times (K+M_t)} = P_0 P_{N \times K} + (1 - P_0) P_{N \times (K+M_t)} \leq P_{mul}^{ors}.$$

∎

Although the adaptive protocol outperforms optimal relay selection, optimal relay selection avoids many implementation issues such as synchronization and carrier frequency offsets. It might also introduce less interference in cellular networks as only one relay transmits at any time.

## E. Numerical Results

Fig. 5 shows $P_c^{ors}$ for a two relay (2,2,4,2) system and $P_c$ for a one relay (2,4,4,2) system, when $\varphi = 20dB$. It can be seen that $P_c^{ors}$ starts decreasing from 1 at SNR $16dB$, which is $6dB$ higher than the equivalent SNR value for $P_c$. Therefore, we expect the two relay system to offer the $4 \times 4$ MIMO DMT for a range of SNR values that is $6dB$ wider than that offered by the single relay system. This advantage is clearly shown in Fig. 6, where the outage probabilities for the two systems are plotted. Note that a much greater performance advantage will occur when $r$ is smaller (i.e., the curve has a steeper slope).

Fig. 7 shows $\{P_{c_i}\}$ in a two relay (2,3,4,1) system for $\varphi = 20dB$, and Fig. 14 shows the system outage probability when $r = 2.3$. Its performance is compared with that of a one relay (2,3,4,1) system. Clearly we observe a significant performance gain for the two relay system. Similarly to the discussion for single relay networks, the system performance is less sensitive to the change of SNR as indicated in *Theorem 2* and *Remark 8*. The network performs the same as a $4 \times 4$ MIMO system within the SNR range where $P_{c_2}$ is large (i.e., $\eta \leq 15dB$). From the SNR value where $P_{c_2}$ decreases to a certain level (e.g., $0.7$), the $4 \times 4$ MIMO DMT no longer dominates the performance and the curve slope changes to that of a $4 \times 3$ MIMO system (i.e., $15dB \leq \eta \leq 25dB$). Finally, when $P_{c_1}$ is sufficiently small, the system performance degrades to that of direct transmission, i.e., a $4 \times 2$ MIMO system. Note that the reason why outage probabilities increase as the SNR increases beyond a certain SNR value, before reaching $1$, is because the transmission rate is higher than the maximal rate that direct transmission can support, i.e., $r > 2$.

## VI. Other Relaying Schemes for Cooperative Multiplexing

In this section we briefly introduce several other relaying schemes that can also result in a multiplexing gain higher than that of direct transmission. Specifically, our discussion covers both full-duplex and half-duplex scenarios.

### A. Full Duplex

*1) Cyclic relaying:* Instead of using a fixed set of antennas $M_t \leq M$ to re-transmit messages during each frame, one might think of using different sets of antennas to re-transmit, in order to obtain a higher diversity gain from the relay to destination link. Based on this idea, we propose a cyclic relaying method, which exploits further the diversity benefits of relaying.

In cyclic relaying, the relay uses $I$ sets of antennas $\mathcal{A}_1, \ldots, \mathcal{A}_I$ to transmit the messages in turn. For the sake of simplicity, we assume that each set $\mathcal{A}_i$ has the same size $M_t$ and $\bigcap_{i=1}^{I} \mathcal{A}_i = \emptyset$. The transmission follows the standard full-duplex protocol described in Section IV except that each message is transmitted in $I$ time slots. Specifically, the source encodes each message $\mathfrak{S}_l$ into a codeword matrix

$$\mathbf{X}_l^{\mathcal{S}} = \left( \begin{array}{ccc} \mathbf{x}_l^{\mathcal{S}(1)} & \cdots & \mathbf{x}_l^{\mathcal{S}(I)} \end{array} \right),$$

where $\mathbf{x}_l^{\mathcal{S}(i)}$ denotes the $i$th $1 \times K$ codeword vector. After the relay receives and decodes $\mathbf{X}_l^{\mathcal{S}}$, it re-encodes the message into a codeword matrix

$$\mathbf{X}_l^{\mathcal{R}} = \left( \begin{array}{ccc} \mathbf{x}_l^{\mathcal{R}(1)} & \cdots & \mathbf{x}_l^{\mathcal{R}(I)} \end{array} \right),$$

where $\mathbf{x}_l^{\mathcal{R}(i)}$ denotes a $1 \times M_t$ codeword vector, and transmit it to the destination. Each $\mathbf{x}_l^{\mathcal{R}(i)}$ is transmitted by the antenna set $\mathcal{A}_i$ at the relay. During the relay transmission of $\mathbf{X}_l^{\mathcal{R}}$, the source simultaneously transmits $\mathbf{X}_{l+1}^{\mathcal{S}}$. Here, we assume that $\mathbf{X}_l^{\mathcal{S}}$ and $\mathbf{X}_l^{\mathcal{R}}$ are generated by two *independent* codebooks. The destination decodes the message $\mathfrak{S}_l$ after receiving it from both the source and the relay, while treating $\mathfrak{S}_{l+1}$ as interference. After $\mathfrak{S}_l$ is decoded, the destination subtracts it from the received signal vector and continues receiving and decoding $\mathfrak{S}_{l+1}$ in the next $I$ time slots.

The system achievable rate, on assuming a perfect source-relay link, can be derived as

$$C = \frac{1}{I} \log \det \left( \mathbf{I} + \eta \mathcal{H}_{\mathcal{S},\mathcal{D}} \mathcal{H}_{\mathcal{S},\mathcal{D}}^H + \eta \mathcal{H}_{\mathcal{R},\mathcal{D}} \mathcal{H}_{\mathcal{R},\mathcal{D}}^H \right) \tag{35}$$

where $\mathcal{H}_{\mathcal{S},\mathcal{D}}$ and $\mathcal{H}_{\mathcal{R},\mathcal{D}}$ denote the block diagonal matrices

$$\mathcal{H}_{\mathcal{S},\mathcal{D}} = \left( \begin{array}{ccc} \mathbf{H}_{\mathcal{S},\mathcal{D}} & & \\ & \ddots & \\ & & \mathbf{H}_{\mathcal{S},\mathcal{D}} \end{array} \right), \mathcal{H}_{\mathcal{R},\mathcal{D}} = \left( \begin{array}{ccc} \mathbf{H}_{\mathcal{R}_{(1)}\mathcal{D}} & & \\ & \ddots & \\ & & \mathbf{H}_{\mathcal{R}_{(I)}\mathcal{D}} \end{array} \right). \tag{36}$$

Here $\mathbf{H}_{\mathcal{R}_{(i)}\mathcal{D}}$ denotes the transfer matrix from the antenna set $\mathcal{A}_i$ to the destination. (35) can be further expressed as

$$C = \frac{1}{I} \sum_{i=1}^{I} \log \det \left( \mathbf{I} + \mathbf{H}_{\mathcal{S}\mathcal{R}_{(i)},\mathcal{D}} \mathbf{H}_{\mathcal{S}\mathcal{R}_{(i)},\mathcal{D}}^H \right). \tag{37}$$

where $\mathbf{H}_{\mathcal{S}\mathcal{R}_{(i)},\mathcal{D}} \overset{\Delta}{=} \left( \begin{array}{cc} \mathbf{H}_{\mathcal{S},\mathcal{D}} & \mathbf{H}_{\mathcal{R}_{(i)},\mathcal{D}} \end{array} \right)$. Note that the elements in $\left\{ \mathbf{H}_{\mathcal{S}\mathcal{R}_{(i)},\mathcal{D}} \right\}$ are correlated, so the exact infinite-SNR DMT curve is difficult to obtain. Instead, using the definition and theory of exchangeable random variables introduced in Section II, we can obtain the following results regarding the bounds for diversity gain and DMT.

*Theorem 3:* On assuming a perfect source-relay channel, we have the following properties for cyclic relaying. (A) The maximal diversity gain for cyclic relaying, *under arbitrary fading* *statistics*, is the same as the maximal diversity gain of an $N \times (K + IM_t)$ MIMO channel. (B) The infinite-SNR DMT for cyclic relaying, *under arbitrary fading statistics*, is lower bounded by that of an $N \times (K + M_t)$ MIMO channel.

*Proof:* See Appendix D. ∎

The adaptive protocol introduced in Section IV.B can be used together with cyclic relaying when the source-relay channel is imperfect. Similar to the analysis in Section IV.B, the system outage probability can be expressed as

$$P_{out}^{cyc} \leq P_c^{cyc} P_{cyc} + (1 - P_c^{cyc}) P_{N \times K}, \tag{38}$$

where $P_{cyc}$ is the outage probability on assuming the perfect source-relay transmission and has the property shown in *Theorem 3* when $\eta \to +\infty$, and where $P_c^{cyc}$ is the probability with which the following constraint holds:

$$\Pi \log \det \left( \mathbf{I} + \varphi \eta \mathbf{H}_{\mathcal{S},\mathcal{R}} \mathbf{H}_{\mathcal{S},\mathcal{R}}^H \right) \geq \log \det \left( \mathbf{I} + \eta \mathcal{H}_{\mathcal{SR},\mathcal{D}} \mathcal{H}_{\mathcal{SR},\mathcal{D}}^H \right), \tag{39}$$

with $\mathcal{H}_{\mathcal{SR},\mathcal{D}} \triangleq \left( \begin{array}{cc} \mathcal{H}_{\mathcal{S},\mathcal{D}} & \mathcal{H}_{\mathcal{R},\mathcal{D}} \end{array} \right)$. When the SNR is high, $P_{cyc}$ offers a diversity gain of nearly $N \times (K + I \times M_t)$ for any fixed transmission rate $R$. This clearly endows cyclic relaying with a performance advantage over fixed relaying.

Following analysis similar to that in Section IV.C, we can observe that cyclic relaying is as *effective* as fixed relaying in the high but finite SNR regime. Specifically, they have the same expression in $\omega$ and both follow the claims in *Lemma 5*. The detailed analysis of these properties is omitted here.

*2) Distributed D-BLAST for a single antenna source:* Thus far, the coding strategies used in the relaying protocols are all based on random coding arguments. In practice, one might expect to design coding schemes that are derived from standard space-time codes or parallel channel codes[9] to approach the performance limits with lower complexity. While general coding design for cooperative multiplexing is an interesting topic beyond the scope of this paper, we consider a single antenna source scenario and show that the existing parallel channel codes can be applied to approach the optimal outage and DMT performance.

---

[9]The DMTs for the schemes proposed thus far may be achieved by parallel channel decoding such as introduced in Section III.B in [10]. However, due to the multiple transmit antennas at each node, specific codes that would achieve the performance limits are still yet to be developed.

Specifically, each transmission frame contains $L$ messages and lasts $L + M_t$ time slots. Each antenna at the relay uses a different (independent) Gaussian codebook to encode the receive message, before transmitting it to the destination. In transmission time slot $i$, the source transmits the $i$th message, which has been encoded into $x_i^{\mathcal{S}}$, while the relay uses its $j$th transmit antenna $\mathcal{R}(j)$ to transmit the $(i - j)$th message, which has been encoded into a codeword $x_{i-j}^{\mathcal{R}(j)}$. The complete encoding and transmission process is illustrated in Table II. Clearly, this structure mimics the D-BLAST structure proposed for point-to-point MIMO systems [24]. We thus call this scheme distributed D-BLAST.

It is well known that the minimum-mean-square-error (MMSE) successive interference cancelation (SIC) algorithm, applied together with parallel channel codes, can achieve the optimal outage and DMT performance for a D-BLAST MIMO system [24]. For the distributed D-BLAST algorithm here, if we apply the parallel channel encoding principle to $x_i^{\mathcal{S}}, x_i^{\mathcal{R}(1)}, \ldots x_i^{\mathcal{R}(M_t)}$, the same conclusion can be drawn.

We note that recently developed approximately universal channel codes for point-to-point MIMO D-BLAST systems, such as the permutation codes using quadrature amplitude modulation (QAM) modulation in [20], can be used in the distributed D-BLAST structure here as well. Thus, cooperative multiplexing gain can *readily* be obtained with a realistic encoding complexity. When the source-relay channel is not perfect, an adaptive protocol can be applied together with distributed D-BLAST to offer the outage probability shown in *Theorem 1*. In a multiple relay scenario, distributed D-BLAST can be applied together with optimal relay selection to offer the better performances shown in *Corollary 2*.

The constraint (19) in *Lemma 5* also applies to the distributed D-BLAST scheme for a single-antenna-source network. One can observe that $\omega = 1/(\mathcal{M}_{\mathcal{SR},\mathcal{D}} - 1) < 1$. Therefore this scheme is *not effective* in the high SNR region. Specifically, we can see that $\eta \lesssim \nu \sqrt[M_t]{\varphi}$ when $M_t < N$. In this scenario, given any value of $\varphi = x(dB)$, the suitable range of SNR is around $\frac{x}{M_t}(dB)$ in order for distributed D-BLAST to perform effectively. Therefore, one might expect it to offer significant benefits for medium SNR values (e.g., $0 \sim 10dB$). Note that direct transmission offers a multiplexing gain of no more than 1 here. Therefore, relaying might still offer significant benefits considering its much higher DMT for moderate SNRs.

## B. Half Duplex: Standard Space-Time Coding

When considering the half duplex setup, we extend the standard space-time random coding scheme proposed by Laneman and Wornell [2] to a multiple antenna scenario. The transmission of each message is divided into two time slots. In the first time slot, the source broadcasts the signal to the relay and the destination. The relay re-encodes the message using an independent codebook and re-transmits it to the destination in the second time slot, while the source remains *silent*.

The input-output relationship, for every pair of transmission time slots, can be expressed as

$$\mathbf{y} = \mathbf{H}\mathbf{x} + \mathbf{n}, \tag{40}$$

where $\mathbf{x} = \left( \begin{array}{cc} \mathbf{x}^{\mathcal{S}} & \mathbf{x}^{\mathcal{R}} \end{array} \right)^{T}$. Here, $\mathbf{x}^{\mathcal{S}}$ is the $1 \times K$ codeword vector at the source, $\mathbf{x}^{\mathcal{R}}$ is the $1 \times M_t$ codeword vector at the relay, and the channel matrix $\mathbf{H}$ can be expressed as

$$\mathbf{H} = \left( \begin{array}{cc} \mathbf{H}_{\mathcal{S},\mathcal{D}} & \\ & \mathbf{H}_{\mathcal{R},\mathcal{D}} \end{array} \right). \tag{41}$$

*Theorem 4:* On assuming a perfect source-relay channel, the infinite-SNR DMT for the standard space-time coding scheme, under Rayleigh fading, is a linear curve connecting the points $(r_i, d_i)$, for $i = 0, 1, \ldots, \min(K, N) + \min(M_t, N)$, where $r_i = \frac{i}{2}$, and $d_i = (K - \alpha_i)(N - \alpha_i) + (M_t - \beta_i)(N - \beta_i)$. The values of $\alpha_i$ and $\beta_i$ can be determined by the logical statement shown below. Specifically, if $K = M_t$, the infinite-SNR DMT is a linear curve connecting the points $(k, 2(K - k)(N - k))$ for integer $k$.

*Proof:* See Appendix D. ∎

Note that *Theorem 4* also applies when the antennas at the relay are distributed, in which case

```
α₀ = β₀ = 0,
for i = 1 : min (K, N) + min (Mₜ, N)
        if (K − αᵢ₋₁) (N − αᵢ₋₁) − (K − αᵢ₋₁ − 1) (N − αᵢ₋₁ − 1) ≥
                    (Mₜ − βᵢ₋₁) (N − βᵢ₋₁) − (Mₜ − βᵢ₋₁ − 1) (N − βᵢ₋₁ − 1)
            αᵢ = αᵢ₋₁ + 1, βᵢ = βᵢ₋₁
        else
            βᵢ = βᵢ₋₁ + 1, αᵢ = αᵢ₋₁
        end
end
```

each antenna should use a different Gaussian codebook to re-encode the message. *Theorem 4* shows that a maximal multiplexing gain of $(\min(K, N) + \min(M_t, N))/2$ can be obtained by using relaying. Compared with the maximal multiplexing gain $\min(K, N)$ achieved by direct transmission, an additional multiplexing gain of $(M_t − K)/2$ can be obtained if $K < M_t \leq N$. We note that this fact is not discovered in [2], which considers only a single antenna network. In fact, it was shown in [2] that the network suffers from a multiplexing loss (i.e, a multiplexing gain of less than 1) when the destination is deployed with only a single antenna, even assuming that the message is correctly decoded at the relays. Therefore, we conclude that the potential of space-time codes is not fully exploited in [2]. In a multi-antenna scenario, random space-time coding in fact not only offers diversity gain, but also multiplexing gain as well. The reason behind this is still the additional degrees of freedom introduced by cooperative encoding, i.e., by concatenating the source and relay codewords and performing joint decoding at the destination.

The adaptive protocol can also be applied to standard space-time coding when the source-relay transmission is imperfect. Following the analysis in Section IV.B, the system outage probability in this scenario can be expressed as

$$P_{out}^{stc} \leq P_c^{stc} P_{stc} + \left(1 − P_c^{stc}\right) P_{dir},  \qquad (42)$$

where $P_{stc}$ is the outage probability on assuming perfect source-relay transmission and has the property shown in *Theorem 4* when $\eta \to +\infty$; $P_{dir}$ is the outage probability when the direct link is used and its infinite SNR DMT is a linear curve connecting the points $(k, (K − 2k)(N − 2k))$;

and $P_c^{stc}$ is the probability with which the following constraint holds:

$$\log \det \left( \mathbf{I} + \varphi \eta \mathbf{H}_{\mathcal{S},\mathcal{R}} \mathbf{H}_{\mathcal{S},\mathcal{R}}^H \right) \geq \log \det \left( \mathbf{I} + \eta \mathbf{H}_{\mathcal{S},\mathcal{D}} \mathbf{H}_{\mathcal{S},\mathcal{D}}^H \right) + \log \det \left( \mathbf{I} + \eta \mathbf{H}_{\mathcal{R},\mathcal{D}} \mathbf{H}_{\mathcal{R},\mathcal{D}}^H \right). \quad (43)$$

When the SNR is high, following the analysis in Section IV. C, it can be seen that the constraints (43) can be expressed in the same form as (16) where

$$\omega = \frac{\min \left( K, M \right)}{\min \left( K, N \right) + \min \left( M_t, N \right) - \min \left( K, M \right)}. \quad (44)$$

Note that at high SNR, we need $K < \min \left( M_t, N \right)$ in order for space time coding to offer a higher multiplexing gain than direct transmission. In this scenario $\omega$ can be re-written as

$$\omega = \frac{K}{\min \left( M_t, N \right)} < 1. \quad (45)$$

Thus relaying is *not effective* at high but finite SNR values when classic space-time coding is applied. Therefore, compared with fixed relaying, space-time coding might require a much higher source to relay path gain $\varphi$ in order to offer multiplexing benefits[10].

## VII. Extensions and Open Issues

In this section, we note some interesting extensions of the results in this paper.

*1) Lognormal shadowing effects:* So far, we have considered the path losses only in terms of large scale fading effects. Lognormal shadowing is also an important characteristic in wireless networks. When the source node is inside a deep lognormal fading area, a relay (or relays) that has good links with both the source and the destination is usually chosen to forward the message. In this case, the effect of the direct link can be ignored compared with those of the source-to-relay and relay-to-destination links. Except for this property, the network models and the analyses in the paper all remain the same in this case. In this sense, lower multiplexing gain might be obtained due to the lack of a direct link. However, relaying can offer a much greater

---

[10]Note that within the range of SNRs where (43) is met with high probability, space-time coding might still offer significant benefits for a small multiplexing gain $r$ or large diversity gain $d$, due to its higher DMT curve for small $r$ as shown in Fig. 2

performance advantage over *direct transmission*, which gives only negligible link reliability and spectral efficiency due to the deep lognormal fading.

*2) Receive cooperation (Compress-and-forward):* Due to the decode-and-forward protocol assumed in this paper, $\varphi$ is usually required to be larger than $0dB$ in most scenarios. When the relays are far away from the source and close to the destination, compress-and-forward becomes a preferred protocol. In this scenario, instead of transmit cooperation, receive cooperation between the relay and the destination is a better choice. Therefore, one might consider a path gain $\varphi$ on the relay to destination link, instead of the source to relay link. The detailed analysis of this situation is an interesting topic for future research.

*3) Transmit and receive cooperation:* One may note that for the system model proposed in this paper, the maximal multiplexing gain is constrained by the antenna array size $N$ at the destination. For the compress-and-forward protocol, the multiplexing gain might be constrained by $K$. Therefore, an interesting question arises as to whether a multiplexing gain that is larger than $\max(K, N)$ can eventually be obtained if we combine transmit cooperation with receive cooperation. In this context, we might use the relays that are close to both the source and the destination. This is another interesting problem for future research.

## VIII. CONCLUSIONS

The paper has offered a new interpretation of the fundamental tradeoff between the rate and outage probability in multi-antenna relay networks, where the concept of DMT is used as a tool to illustrate such a tradeoff and to measure the system performance. Specifically, several protocols have been proposed for decode-and-forward relaying. It has been shown that these schemes can offer significant performance advantages over the direct link in terms of multiplexing gain. The main results of the paper are summarized in Table III for single relay networks, and Table IV for multiple relay networks. Unlike most previous work in this area, which concentrates only on *cooperative diversity*, this work has opened a new direction for exploiting *cooperative multiplexing* in wireless relay networks.

APPENDIX

## A. Proof of Lemma 4

The RHS of (6) can be written as $\det\left(\mathbf{I} + \left(\begin{array}{cc}\mathbf{U} & \mathbf{V}\end{array}\right)\left(\begin{array}{cc}\mathbf{U} & \mathbf{V}\end{array}\right)^H\right)$, where $\mathbf{U} = [\mathbf{u}_1\mathbf{u}_2\cdots\mathbf{u}_{m_1}]$ and $\mathbf{V} = [\mathbf{v}_1\mathbf{v}_2\cdots\mathbf{v}_{m_2}]$. Using the fact that $\det(\mathbf{I} + \mathbf{AB}) = \det(\mathbf{I} + \mathbf{BA})$, we have

$$\det\left(\mathbf{I} + \left(\begin{array}{cc}\mathbf{U} & \mathbf{V}\end{array}\right)\left(\begin{array}{cc}\mathbf{U} & \mathbf{V}\end{array}\right)^H\right) = \det\left(\begin{array}{cc}\mathbf{I} + \mathbf{U}^H\mathbf{U} & \mathbf{U}^H\mathbf{V} \\ \mathbf{V}^H\mathbf{U} & \mathbf{I} + \mathbf{V}^H\mathbf{V}\end{array}\right).$$

Using Fischer's inequality, we have

$$\begin{aligned}\det\left(\begin{array}{cc}\mathbf{I} + \mathbf{U}^H\mathbf{U} & \mathbf{U}^H\mathbf{V} \\ \mathbf{V}^H\mathbf{U} & \mathbf{I} + \mathbf{V}^H\mathbf{V}\end{array}\right) &\leqslant \det\left(\mathbf{I} + \mathbf{U}^H\mathbf{U}\right)\det\left(\mathbf{I} + \mathbf{V}^H\mathbf{V}\right) \\ &= \det\left(\mathbf{I} + \mathbf{U}\mathbf{U}^H\right)\det\left(\mathbf{I} + \mathbf{V}\mathbf{V}^H\right) \\ &= \det\left(\mathbf{I} + \sum\mathbf{u}_i\mathbf{u}_i^H\right)\left(\mathbf{I} + \sum\mathbf{v}_i\mathbf{v}_i^H\right),\end{aligned}$$

and the lemma follows.

## B. Proof of Theorem 1

The proof can be demonstrated through using the formula

$$P_{out}^{adp} = \Pr\left(C_{\mathcal{S},\mathcal{R}} \geq C_{\mathcal{S}\mathcal{R},\mathcal{D}}\right)P_{out}^{adp|C_{\mathcal{S},\mathcal{R}} \geq C_{\mathcal{S}\mathcal{R},\mathcal{D}}} + \left(1 - \Pr\left(C_{\mathcal{S},\mathcal{R}} \geq C_{\mathcal{S}\mathcal{R},\mathcal{D}}\right)\right)P_{out}^{adp|C_{\mathcal{S},\mathcal{R}} < C_{\mathcal{S}\mathcal{R},\mathcal{D}}} \quad (46)$$

where $P_{out}^{adp|C_{\mathcal{S},\mathcal{R}} \geq C_{\mathcal{S}\mathcal{R},\mathcal{D}}}$ $\left(P_{out}^{adp|C_{\mathcal{S},\mathcal{R}} < C_{\mathcal{S}\mathcal{R},\mathcal{D}}}\right)$ is the outage probability given that $C_{\mathcal{S},\mathcal{R}} \geq C_{\mathcal{S}\mathcal{R},\mathcal{D}}$ $\left(C_{\mathcal{S},\mathcal{R}} < C_{\mathcal{S}\mathcal{R},\mathcal{D}}\right)$, and $C_{\mathcal{S}\mathcal{R},\mathcal{D}}$ can be written as

$$C_{\mathcal{S}\mathcal{R},\mathcal{D}} = \log\det\left(\mathbf{I} + \eta\mathbf{H}_{\mathcal{S}\mathcal{R},\mathcal{D}}\mathbf{H}_{\mathcal{S}\mathcal{R},\mathcal{D}}^H\right). \quad (47)$$

 

Further, we obtain

$$
\begin{aligned}
P_{out}^{adp|C_{\mathcal{S},\mathcal{R}} \geq C_{\mathcal{S}\mathcal{R},\mathcal{D}}} \leq {} & \Pr\left(C_{\mathcal{S},\mathcal{R}} < R | C_{\mathcal{S},\mathcal{R}} \geq C_{\mathcal{S}\mathcal{R},\mathcal{D}}\right) + \\
& \Pr\left(C_{\mathcal{S},\mathcal{R}} > R | C_{\mathcal{S},\mathcal{R}} \geq C_{\mathcal{S}\mathcal{R},\mathcal{D}}\right) \Pr\left(C_{\mathcal{S}\mathcal{R},\mathcal{D}} < R | C_{\mathcal{S},\mathcal{R}} > R, C_{\mathcal{S},\mathcal{R}} \geq C_{\mathcal{S}\mathcal{R},\mathcal{D}}\right) \\
= {} & \Pr\left(C_{\mathcal{S},\mathcal{R}} < R \ \ \text{or} \ \ C_{\mathcal{S}\mathcal{R},\mathcal{D}} < R | C_{\mathcal{S},\mathcal{R}} \geq C_{\mathcal{S}\mathcal{R},\mathcal{D}}\right) \\
= {} & \Pr\left(C_{\mathcal{S}\mathcal{R},\mathcal{D}} < R | C_{\mathcal{S},\mathcal{R}} \geq C_{\mathcal{S}\mathcal{R},\mathcal{D}}\right) + \\
& \Pr\left(C_{\mathcal{S}\mathcal{R},\mathcal{D}} > R | C_{\mathcal{S},\mathcal{R}} \geq C_{\mathcal{S}\mathcal{R},\mathcal{D}}\right) \Pr\left(C_{\mathcal{S},\mathcal{R}} < R | C_{\mathcal{S}\mathcal{R},\mathcal{D}} > R, C_{\mathcal{S},\mathcal{R}} \geq C_{\mathcal{S}\mathcal{R},\mathcal{D}}\right),
\end{aligned}
$$

since

$$
\Pr\left(C_{\mathcal{S},\mathcal{R}} < R | C_{\mathcal{S}\mathcal{R},\mathcal{D}} > R, C_{\mathcal{S},\mathcal{R}} \geq C_{\mathcal{S}\mathcal{R},\mathcal{D}}\right) = 0.
$$

Therefore, we obtain

$$
P_{out}^{adp|C_{\mathcal{S},\mathcal{R}} \geq C_{\mathcal{S}\mathcal{R},\mathcal{D}}} \leq \Pr\left(C_{\mathcal{S}\mathcal{R},\mathcal{D}} < R | C_{\mathcal{S},\mathcal{R}} \geq C_{\mathcal{S}\mathcal{R},\mathcal{D}}\right) = \Pr\left(C_{\mathcal{S}\mathcal{R},\mathcal{D}} < R\right)
$$

and thus complete the first part of the proof.

Next, we write

$$
\begin{aligned}
P_{out}^{adp|C_{\mathcal{S},\mathcal{R}} < C_{\mathcal{S}\mathcal{R},\mathcal{D}}} = {} & \Pr\left(C_{\mathcal{S},\mathcal{D}} < R | C_{\mathcal{S},\mathcal{R}} < C_{\mathcal{S}\mathcal{R},\mathcal{D}}\right) \Pr\left(C_{\mathcal{S},\mathcal{R}} < R | C_{\mathcal{S},\mathcal{R}} < C_{\mathcal{S}\mathcal{R},\mathcal{D}}\right) + \\
& \Pr\left(C_{\mathcal{S},\mathcal{R}} > R | C_{\mathcal{S},\mathcal{R}} < C_{\mathcal{S}\mathcal{R},\mathcal{D}}\right) \Pr\left(C_{\mathcal{S}\mathcal{R},\mathcal{D}} < R | C_{\mathcal{S},\mathcal{R}} > R, C_{\mathcal{S},\mathcal{R}} < C_{\mathcal{S}\mathcal{R},\mathcal{D}}\right)
\end{aligned}
$$

Also note that $\Pr\left(C_{\mathcal{S}\mathcal{R},\mathcal{D}} < R | C_{\mathcal{S},\mathcal{R}} > R, C_{\mathcal{S},\mathcal{R}} < C_{\mathcal{S}\mathcal{R},\mathcal{D}}\right) = 0$, so that we obtain

$$
\begin{aligned}
P_{out}^{adp|C_{\mathcal{S},\mathcal{R}} < C_{\mathcal{S}\mathcal{R},\mathcal{D}}} = {} & \Pr\left(C_{\mathcal{S},\mathcal{D}} < R | C_{\mathcal{S},\mathcal{R}} < C_{\mathcal{S}\mathcal{R},\mathcal{D}}\right) \Pr\left(C_{\mathcal{S},\mathcal{R}} < R | C_{\mathcal{S},\mathcal{R}} < C_{\mathcal{S}\mathcal{R},\mathcal{D}}\right) \\
\leq {} & \Pr\left(C_{\mathcal{S},\mathcal{D}} < R\right).
\end{aligned} \tag{48}
$$

Therefore we obtain an upper bound

$$
P_{out}^{adp} \leq \Pr\left(C_{\mathcal{S},\mathcal{R}} \geq C_{\mathcal{S}\mathcal{R},\mathcal{D}}\right) \Pr\left(C_{\mathcal{S}\mathcal{R},\mathcal{D}} < R\right) + \left(1 - \Pr\left(C_{\mathcal{S},\mathcal{R}} \geq C_{\mathcal{S}\mathcal{R},\mathcal{D}}\right)\right) \Pr\left(C_{\mathcal{S},\mathcal{D}} < R\right). \tag{49}
$$

Since $\Pr\left(C_{\mathcal{S}\mathcal{R},\mathcal{D}} < R\right) = P_{N \times (K + M_t)}$, $\Pr\left(C_{\mathcal{S},\mathcal{D}} < R\right) = P_{N \times K}$, and $\Pr\left(C_{\mathcal{S},\mathcal{R}} \geq C_{\mathcal{S}\mathcal{R},\mathcal{D}}\right) = P_c$,

the proof is thus complete.

### C. Proof of Theorem 2

First, we define the events $\mathcal{X}_0, \mathcal{X}_1, \cdots, \mathcal{X}_{\mathfrak{R}-1}$ as follows:

$$
\begin{aligned}
\mathcal{X}_0 &= \left\{ C_{\mathcal{S}, \mathcal{R}_j} < C_{\mathcal{SO}_1, D}, \exists j \in \mathcal{O}_1, \forall \mathcal{O}_1 \subseteq \{\mathcal{R}_k\} \right\}, \\
\mathcal{X}_i &= \left\{ C_{\mathcal{S}, \mathcal{R}_j} < C_{\mathcal{SO}_{i+1}, \mathcal{D}}, \exists j \in \mathcal{O}_{i+1}, \forall \mathcal{O}_{i+1} \subseteq \{\mathcal{R}_k\} \right\}, i = 1, \ldots, \mathfrak{R}-1.
\end{aligned}
$$

Also, define the events $\mathcal{Y}_0, \mathcal{Y}_1, \cdots, \mathcal{Y}_{\mathfrak{R}}$ as follows:

$$
\begin{aligned}
\mathcal{Y}_0 &= \mathcal{X}_0, \\
\mathcal{Y}_i &= \bar{\mathcal{X}}_{i-1} \cap \mathcal{X}_i, \\
\mathcal{Y}_{\mathfrak{R}} &= \bar{\mathcal{X}}_{\mathfrak{R}-1}.
\end{aligned}
$$

It can be shown that any event that belongs to $\mathcal{X}_i$ also belongs to $\mathcal{X}_{i+1}$. The proof of this statement can be easily obtained by using *Lemma 2(d)* and is thus omitted here. Therefore, $\mathcal{X}_i \subseteq \mathcal{X}_{i+1}$. It can also be observed that

$$
\mathcal{X}_i \cup \mathcal{Y}_{i+1} = \mathcal{X}_i \cup \bar{\mathcal{X}}_i \cap \mathcal{X}_{i+1} = \mathcal{X}_{i+1}.
$$

and

$$
\mathcal{X}_i \cap \mathcal{Y}_{i+1} = \mathcal{X}_i \cap \bar{\mathcal{X}}_i \cap \mathcal{X}_{i+1} = \emptyset.
$$

We can thus conclude that

$$
\Pr\left(\mathcal{Y}_{i+1}\right) = \Pr\left(\mathcal{X}_{i+1}\right) - \Pr\left(\mathcal{X}_i\right) \tag{50}
$$

for $i = 1, \ldots, \mathfrak{R}-2$, and $\Pr\left(\mathcal{Y}_{\mathfrak{R}}\right) = 1 - \Pr\left(\mathcal{X}_{\mathfrak{R}-1}\right)$.

Note that each element $P_i$ ($i = 1, \ldots, \mathfrak{R}$) in (31) can be re-written as

$$
P_i = \Pr\left(\mathcal{O}_i | C_{\mathcal{S}, \mathcal{R}_j} \geq R, \forall j \in \mathcal{O}_i, \text{and } C_{\mathcal{S}, \mathcal{R}_j} < R, \forall j \in \bar{\mathcal{O}}_i\right) \triangleq \Pr\left(\mathcal{Z}_i\right). \tag{51}
$$

Then, each element $P_i P_{N \times (K+iM_t)}$ $(i = 1, \ldots, \Re - 1)$ in (31) can be re-written as

$$
\begin{aligned}
P_i P_{N \times (K+iM_t)} &= \Pr\left(\mathcal{Z}_i\right) \Pr\left(C_{\mathcal{SO}_i, \mathcal{D}} < R | \mathcal{Z}_i\right) \\
&= \Pr\left(\mathcal{X}_{i-1} | \mathcal{X}_i\right) \Pr\left(\mathcal{Z}_i | \mathcal{X}_{i-1}, \mathcal{X}_i\right) \Pr\left(C_{\mathcal{SO}_i, \mathcal{D}} < R | \mathcal{Z}_i, \mathcal{X}_{i-1}, \mathcal{X}_i\right) + \\
&\quad \Pr\left(\bar{\mathcal{X}}_{i-1} | \mathcal{X}_i\right) \Pr\left(\mathcal{Z}_i | \bar{\mathcal{X}}_{i-1}, \mathcal{X}_i\right) \Pr\left(C_{\mathcal{SO}_i, \mathcal{D}} < R | \mathcal{Z}_i, \bar{\mathcal{X}}_{i-1}, \mathcal{X}_i\right).
\end{aligned}
\tag{52}
$$

In fact, it can be seen that $\Pr\left(C_{\mathcal{SO}_i, \mathcal{D}} < R | \mathcal{Z}_i, \mathcal{X}_{i-1}\right) = 0$. Therefore, the first term in (52) can be ignored. Thus we obtain

$$
P_i P_{N \times (K+iM_t)} \leq \Pr\left(\bar{\mathcal{X}}_{i-1} | \mathcal{X}_i\right) \Pr\left(C_{\mathcal{SO}_i, \mathcal{D}} < R\right).
\tag{53}
$$

From (50), we obtain

$$
\Pr\left(\bar{\mathcal{X}}_{i-1} | \mathcal{X}_i\right) = \frac{\Pr\left(\bar{\mathcal{X}}_{i-1} \cap \mathcal{X}_i\right)}{\Pr\left(\mathcal{X}_i\right)} = \frac{\Pr\left(\mathcal{Y}_i\right)}{\Pr\left(\mathcal{X}_i\right)} = \frac{\Pr\left(\mathcal{X}_i\right) - \Pr\left(\mathcal{X}_{i-1}\right)}{\Pr\left(\mathcal{X}_i\right)}.
\tag{54}
$$

For $i = \Re$, we have

$$
P_\Re P_{N \times (K+\Re M_t)} \leq \Pr\left(\bar{\mathcal{X}}_{i-1}\right) \Pr\left(C_{\mathcal{SO}_i, \Re} < R\right) = \left(1 - \Pr\left(\mathcal{X}_{i-1}\right)\right) \Pr\left(\mathcal{C}_{\mathcal{SO}_i, \Re} < R\right).
\tag{55}
$$

On noting that $P_{c_i}^{\emptyset} = \Pr\left(\mathcal{X}_{i-1}\right)$, the proof is thus complete.

### D. Proof of Theorem 3

Note that the elements in $\left\{\mathbf{H}_{\mathcal{SR}_{(i)}, \mathcal{D}}\right\}$ form an *exchangeable sequence*. The outage probability can be thus characterized as

$$
\begin{aligned}
P_{out} &\leq \Pr\left(\sum_{i=1}^{I} \log \det\left(\mathbf{I} + \mathbf{H}_{\mathcal{SR}_{(i)}, \mathcal{D}} \mathbf{H}_{\mathcal{SR}_{(i)}, \mathcal{D}}^{H}\right) < IR\right) \tag{56} \\
&\leq \Pr\left(I \times \min\left\{\log \det\left(\mathbf{I} + \mathbf{H}_{\mathcal{SR}_{(i)}, \mathcal{D}} \mathbf{H}_{\mathcal{SR}_{(i)}, \mathcal{D}}^{H}\right)\right\} < IR\right) \tag{57} \\
&< 1 - \left(1 - P_{N \times (K+M_t)}\right)^{I}, \tag{58}
\end{aligned}
$$

where we used *Lemma 1* to obtain (58) from (57). Using the inequality $mx > 1 - (1-x)^m$ for $0 < x < 1$ and integral $m$, the diversity gain for the system can be lower bounded by

$$d_{cyc} = -\lim_{\eta \to +\infty} \frac{\log P_{out}}{\log \eta} \geq -\lim_{\eta \to +\infty} \frac{\log m P_{N \times (K+M_t)}}{\log \eta} = d_{N \times (K+M_t)}. \tag{59}$$

To obtain the maximal value for $d_{cyc}$, we focus on the probability bound (56) for a fixed rate $R$ (i.e., $r = 0$). Using *Lemma 4*, we obtain

$$\sum_{i=1}^{I} \log \det \left( \mathbf{I} + \mathbf{H}_{\mathcal{SR}_{(i)},\mathcal{D}} \mathbf{H}_{\mathcal{SR}_{(i)},\mathcal{D}}^H \right) \geq \log \det \left( \mathbf{I} + \sum_{i=1}^{I} \mathbf{H}_{\mathcal{SR}_{(i)},\mathcal{D}} \mathbf{H}_{\mathcal{SR}_{(i)},\mathcal{D}}^H \right) \tag{60}$$

$$= \log \det \left( \mathbf{I} + I \mathbf{H}_{\mathcal{S},\mathcal{D}} \mathbf{H}_{\mathcal{S},\mathcal{D}}^H + \sum_{i=1}^{I} \mathbf{H}_{\mathcal{R}_{(i)},\mathcal{D}} \mathbf{H}_{\mathcal{R}_{(i)},\mathcal{D}}^H \right) \tag{61}$$

$$\geq \log \det \left( \mathbf{I} + \mathbf{H}_{\mathcal{S},\mathcal{D}} \mathbf{H}_{\mathcal{S},\mathcal{D}}^H + \sum_{i=1}^{I} \mathbf{H}_{\mathcal{R}_{(i)},\mathcal{D}} \mathbf{H}_{\mathcal{R}_{(i)},\mathcal{D}}^H \right) \tag{62}$$

$$= \log \det \left( \mathbf{I} + \tilde{\mathbf{H}}_{\mathcal{SR},\mathcal{D}} \tilde{\mathbf{H}}_{\mathcal{SR},\mathcal{D}}^H \right) \tag{63}$$

where we have used *Lemma 2(a)* and *Lemma 2(d)* to move from (61) to (62). The matrix $\tilde{\mathbf{H}}_{SR,D}$ can be expressed as $\left( \begin{array}{cccc} \mathbf{H}_{\mathcal{S},\mathcal{D}} & \mathbf{H}_{\mathcal{R}_{(1)},\mathcal{D}} & \cdots & \mathbf{H}_{\mathcal{R}_{(I)},\mathcal{D}} \end{array} \right)$, and it can be considered to be equivalent to a matrix for an $N \times (K + IM_t)$ MIMO channel that has a maximal diversity gain of $N \times (K + IM_t)$ when $r = 0$. Therefore, we can conclude that the maximal $d_{cyc}$ cannot be worse than $N \times (K + IM_t)$. The same upper bound for maximal $d_{cyc}$ can be obtained by showing

$$\sum_{i=1}^{I} \log \det \left( \mathbf{I} + \mathbf{H}_{\mathcal{SR}_{(i)},\mathcal{D}} \mathbf{H}_{\mathcal{SR}_{(i)},\mathcal{D}}^H \right) \leq I \log \det \left( \mathbf{I} + \mathbf{H}_{\mathcal{SR},\mathcal{D}} \tilde{\mathbf{H}}_{\mathcal{SR},\mathcal{D}}^H \right), \tag{64}$$

and the proof is complete.

*E. Proof of Theorem 4*

We use the method in [3] to prove the theorem. On assuming that the average transmission rate changes as $\bar{R} = \frac{1}{2}R = r \log \eta$, the outage probability can be expressed as

$$
\begin{aligned}
P_{out} &= \Pr\left(\log \det\left(\mathbf{I} + \eta \mathbf{H}_{\mathcal{S},\mathcal{D}} \mathbf{H}_{\mathcal{S},\mathcal{D}}^H\right) \det\left(\mathbf{I} + \eta \mathbf{H}_{\mathcal{R},\mathcal{D}} \mathbf{H}_{\mathcal{R},\mathcal{D}}^H\right) < 2R\right) \\
&= \Pr\left(\prod_i \left(1 + \eta \lambda_i^{\mathcal{S},\mathcal{D}}\right) \prod_j \left(1 + \eta \lambda_j^{\mathcal{R},\mathcal{D}}\right) < \eta^{2r}\right),
\end{aligned} \tag{65}
$$

where $\left\{\lambda_i^{\mathcal{S},\mathcal{D}}\right\}$ $\left(\left\{\lambda_j^{\mathcal{R},\mathcal{D}}\right\}\right)$ denote the eigenvalues of $\mathbf{H}_{\mathcal{S},\mathcal{D}} \mathbf{H}_{\mathcal{S},\mathcal{D}}^H$ $\left(\mathbf{H}_{\mathcal{R},\mathcal{D}} \mathbf{H}_{\mathcal{R},\mathcal{D}}^H\right)$, with $\lambda_1^{\mathcal{S},\mathcal{D}} \leq \lambda_2^{\mathcal{S},\mathcal{D}} \leq \cdots \leq \lambda_{\min(K,N)}^{\mathcal{S},\mathcal{D}}$. Let $\lambda_i^{\mathcal{R},\mathcal{D}} \doteq \eta^{-\alpha_i}$ and $\lambda_i^{\mathcal{R},\mathcal{D}} \doteq \eta^{-\beta_i}$. At high SNR, the preceding expression can be further approximated as

$$
P_{out} \doteq \Pr\left(\sum_i \left(1 - \alpha_i\right)^+ + \sum_j \left(1 - \beta_j\right)^+ < 2r\right). \tag{66}
$$

Note that the following property holds for the probability density functions (PDFs) of $\{\alpha_i\}$ [3]:

$$
p_{(\alpha_1, \cdots, \alpha_n)} \doteq \begin{cases} 0, & \text{for } \alpha_i < 0 \\ \eta^{\sum\limits_{i=1}^{\min(K,N)} -(2i-1+|K-N|)\alpha_i}, & \text{for } \alpha_i \geq 0 \end{cases}. \tag{67}
$$

A similar principle also applies to $\{\beta_i\}$. Also note the fact that the two vectors $\alpha = (\alpha_1, \cdots, \alpha_n)$ and $\beta = (\beta_1, \cdots, \beta_n)$ are statistically independent of each other. Therefore, the outage probability for high SNR can be characterized as

$$
P_{out} \doteq \int_{O^+} \eta^{-\left(\sum\limits_{i=1}^{\min(K,N)} (2i-1+|K-N|)\alpha_i + \sum\limits_{j=1}^{\min(M,N)} (2j-1+|M-N|)\beta_j\right)} do, \tag{68}
$$

where $o \overset{\Delta}{=} (\alpha \ \ \beta)$ and $O^+$ denotes the set of outage events, which can be expressed as

$$
O^+ = \Bigg\{ o \in R^{(\min(K,N)+\min(M,N))+} | \alpha_1 \geq \cdots \geq \alpha_{\min(K,N)} \geq 0, \beta_1 \geq \cdots \geq \beta_{\min(K,N)} \geq 0,
$$
$$
\text{and} \sum_i \left(1 - \alpha_i\right)^+ + \sum_j \left(1 - \beta_j\right)^+ < 2r \Bigg\} \tag{69}
$$

                                                                                      

The diversity gain thus can be calculated as

$$d\left(r\right) = \inf_{o \in O^+} \left( \sum_{i=1}^{\min(K,N)} \left(2i - 1 + |K - N|\right) \alpha_i + \sum_{j=1}^{\min(M,N)} \left(2j - 1 + |M - N|\right) \beta_j \right). \quad (70)$$

The value of $d(r)$ can be calculated explicitly for each value of $r$. Specifically, for $\alpha_i = 1$ ($i = 1, \ldots, \min(K, N) - k$) and $\alpha_j = 0$ ($j = \min(K, N) - k + 1, \ldots, \min(K, N)$), We have

$$\sum_{i=1}^{\min(K,N)} \left(2i - 1 + |K - N|\right) \alpha_i = (K - k)(N - k). \quad (71)$$

For $r = 0$, all $\alpha_i = 0$ , all $\beta_i = 0$, and $d(r)$ obtains its largest value $NK + MK$. When $r$ is increased by $1/2$, in order to obtain the minimal value $d(r)$, the element (either in $\{\alpha_i\}$ or $\{\beta_i\}$) that corresponds to the largest component in the summation (70) should be changed from one to zero. When $r = 1/2$, clearly the element is $\alpha_{\min(N,K)}$ if

$$2\min(N, K) - 1 + |K - N| > 2\min(M, K) - 1 + |K - M|, \quad (72)$$

else it is $\beta_{\min(N,K)}$. Because of (71), inequality (72) can be re-expressed as changing $\alpha_{\min(N,K)}$ to zero if

$$KN - (K - 1)(N - 1) > MN - (M - 1)(N - 1). \quad (73)$$

$d(r)$ in this scenario can be expressed as

$$d(r) = (K - 1)(N - 1) + MN. \quad (74)$$

Furthermore, it can be observed that the curve $d(r)$ is linear between $0 < r < 1/2$. The same process continues each time when $r$ is increased by $1/2$, and finally stops at $r = \frac{\min(K,N) + \min(M,N)}{2}$, where all $\alpha_i$ and $\beta_i$ are equal to zero and $d(r) = 0$. More specifically, the calculation can be evaluated using the pseudo-code in the text box below *Theorem 4*.

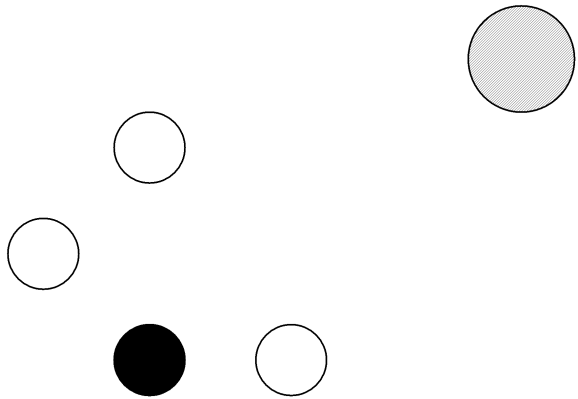

Fig. 1. System model for a single source network. The black node is the source, the white nodes are the relays, and the grey node is the destination.

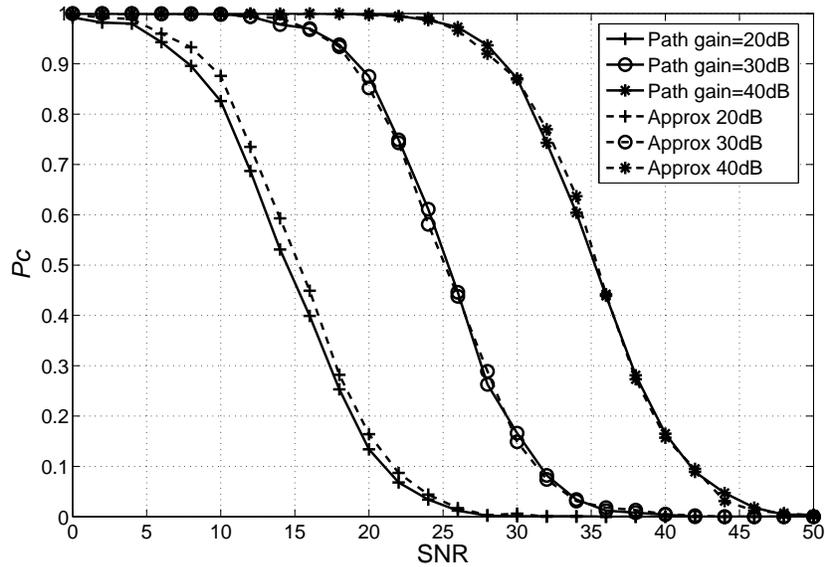

Fig. 2. $P_c$ for different values of $\varphi$ for (2,2,4,2), $\omega = 1$. The dashed curves are the approximations of $P_c$, i.e., $P_\nu$.

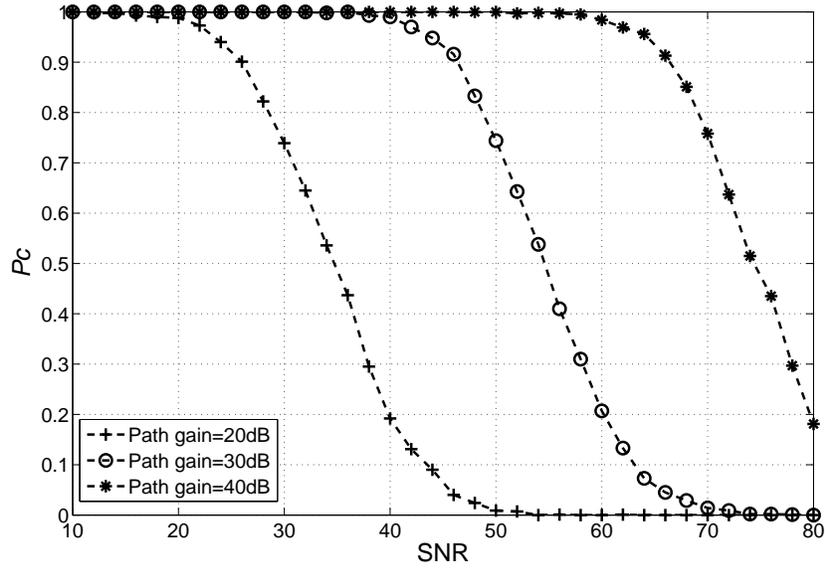

Fig. 3. $P_c$ for different values of $\varphi$ for (2,3,4,1), $\omega = 2$.

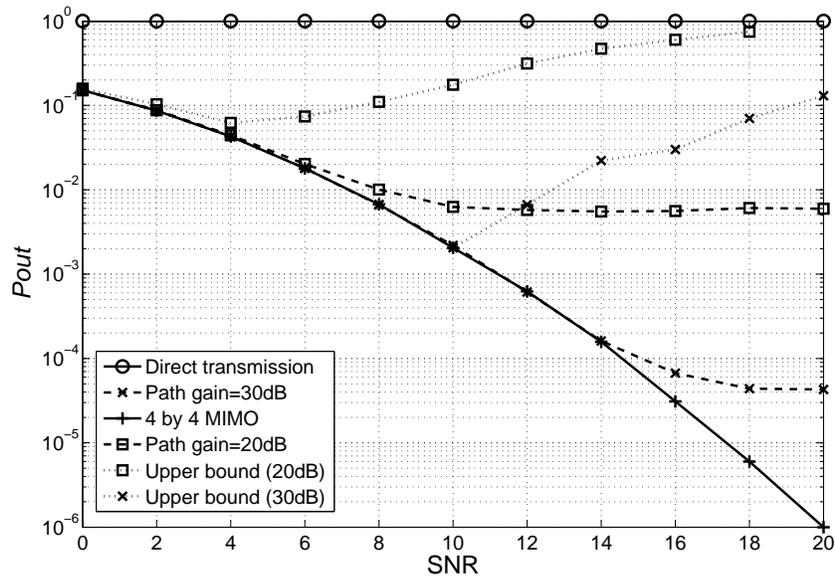

Fig. 4. $P_{out}^{adp}$ for different values of $\varphi$ for (2,2,4,2), when $r = 2$. The solid curve with marks "o" depicts direct transmission (i.e., a $2 \times 2$ MIMO channel). The solid curve with marks "+" depicts a $4 \times 4$ MIMO channel. The dashed and dotted curves with square marks depict $P_{out}^{adp}$ for $\varphi = 20dB$ and its upper bound. The dashed and dotted curves with square marks depict $P_{out}^{adp}$ for $\varphi = 30dB$ and its upper bound.

                                                                

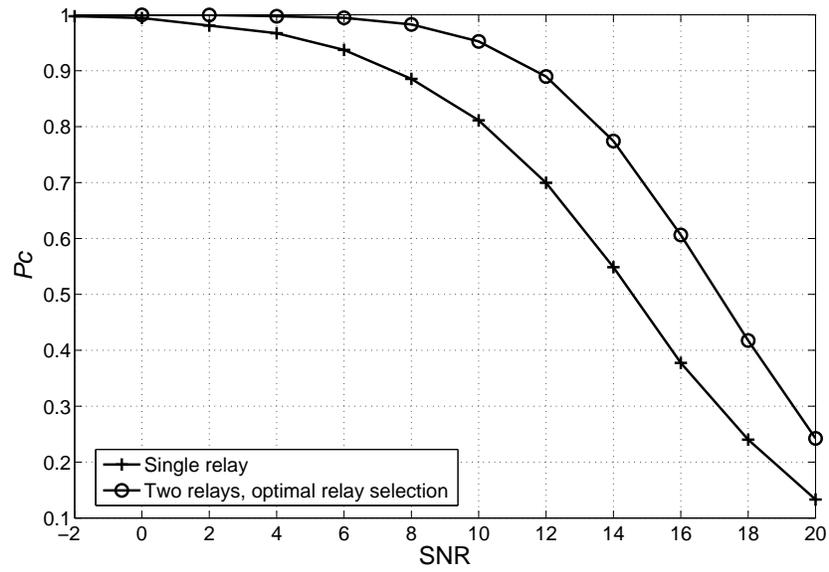

Fig. 5. $p_c^{ors}$ for a two relay (2,2,4,2) system and $p_c$ for a one relay (2,2,4,2) system, when $\varphi = 20dB$.

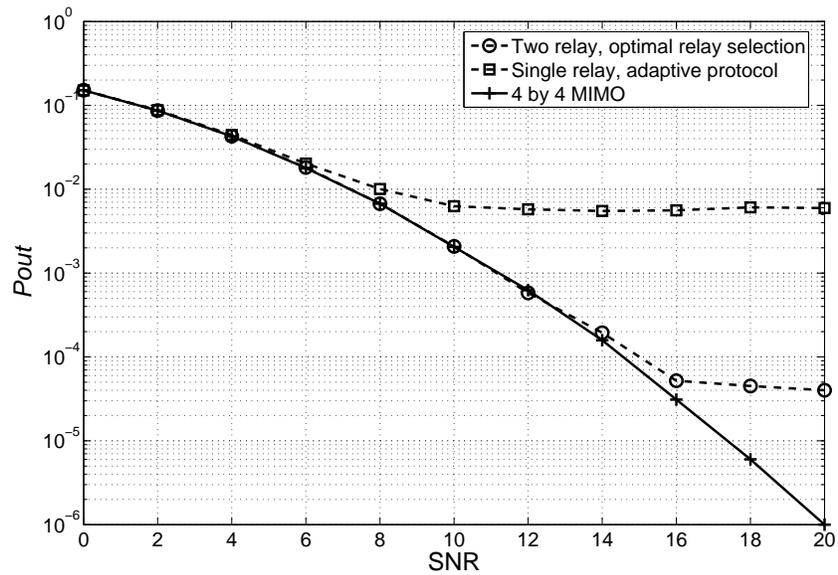

Fig. 6. The outage probabilities for a one relay (2,2,4,2) system and a two relay (2,2,4,2) system when optimal relay selection is applied. $\varphi = 20dB$.

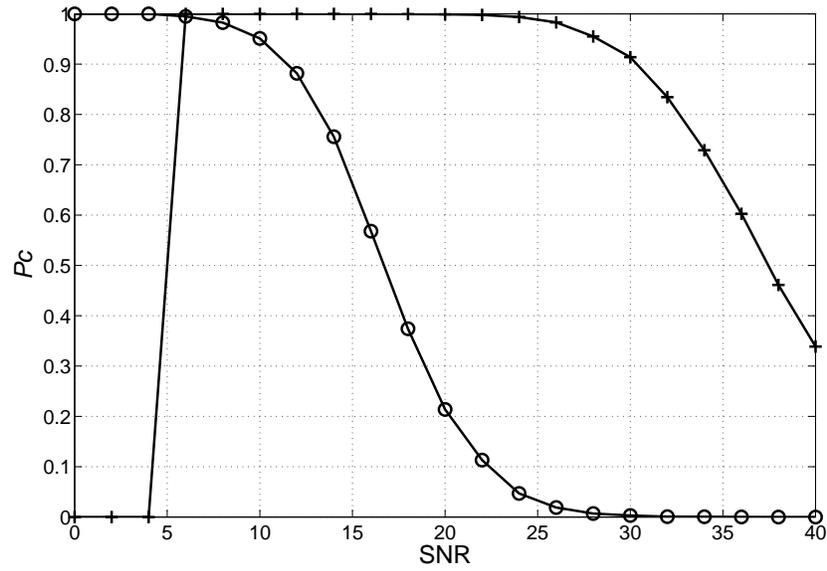

Fig. 7. $P_{c_i}$ in a two relay (2,3,4,1) system using adaptive protocols, when $\varphi = 20dB$. The curve with marks "o" shows $P_{c_2}$. The curve with marks "+" shows $P_{c_1}$.

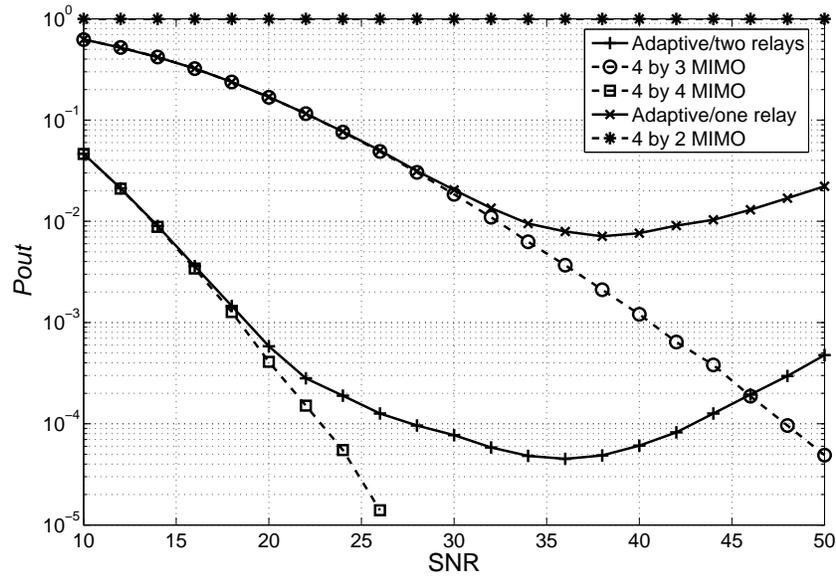

Fig. 8. Outage probability for fixed relaying in a two relay (2,3,4,1) system using adaptive protocols, when $\varphi = 20dB$ and $r = 2.3$.

| $\mathcal{S}$ | $\mathbf{x}_1^{\mathcal{S}(1)}$ | $\mathbf{x}_1^{\mathcal{S}(2)}$ | $\mathbf{x}_2^{\mathcal{S}(1)}$ | $\mathbf{x}_2^{\mathcal{S}(2)}$ | | |
|---|---|---|---|---|---|---|
| $\mathcal{A}_1$ | | | $\mathbf{x}_1^{\mathcal{R}(1)}$ | | $\mathbf{x}_2^{\mathcal{R}(1)}$ | |
| $\mathcal{A}_2$ | | | | $\mathbf{x}_1^{\mathcal{R}(2)}$ | | $\mathbf{x}_2^{\mathcal{R}(2)}$ |

TABLE I

TRANSMISSION SCHEDULE FOR CYCLIC RELAYING FOR $I = 2, L = 2$, WHERE $\mathcal{A}_i$ ARE THE ANTENNA SETS AT THE RELAY.

| $\mathcal{S}$ | $x_1^{\mathcal{S}}$ | $x_2^{\mathcal{S}}$ | $x_3^{\mathcal{S}}$ | ... | $x_L^{\mathcal{S}}$ |
|---|---|---|---|---|---|
| $\mathcal{R}(1)$ | | $x_1^{\mathcal{R}(1)}$ | $x_2^{\mathcal{R}(1)}$ | $x_3^{\mathcal{R}(1)}$ | ... |
| $\mathcal{R}(2)$ | | | $x_1^{\mathcal{R}(2)}$ | $x_2^{\mathcal{R}(2)}$ | $x_3^{\mathcal{R}(2)}$ |
| ... | | | | ... | ... |
| $\mathcal{R}(M_t)$ | | | | | $x_1^{\mathcal{R}(M_t)}$ |

TABLE II

RELAYING SCHEME FOR A SINGLE ANTENNA SOURCE.

| | Outage probability | Conditions for relaying to be effective |
|---|---|---|
| Fixed relaying | See (12) | See (19) |
| Cyclic relaying | See (38) | See (19) |
| Distributed D-BLAST | See (12) | Not effective |
| Standard space-time coding | See (42) | Not effective |

TABLE III

A SUMMARY OF THE PERFORMANCE GIVEN BY DIFFERENT RELAYING SCHEMES IN A SINGLE RELAY SCENARIO.

|  | Assume $\Re$ relays available and each relay has $M_t$ transmit antennas and $M$ antennas in total. |
|---|---|
| Optimal relay selection | $d \to d_{N \times (K+M_t)}$ as $\Re \to +\infty$ |
| Adaptive protocol | Always performs better than optimal relay selection, As SNR increases from $-\infty$ to $+\infty$, the DMT changes as $d_{N \times (K+\Re M_t)} \to d_{N \times (K+(\Re-1)M_t)} \to \cdots \to d_{N \times K}$ |

TABLE IV

A SUMMARY OF THE PERFORMANCE GIVEN BY DIFFERENT PROTOCOLS IN A MULTIPLE RELAY SCENARIO, IN WHICH FIXED RELAYING IS APPLIED.

**Yijia (Richard) Fan** received his BEng degree in electrical engineering from Shanghai Jiao Tong University (SJTU), Shanghai, P.R. China, in July 2003, and PhD degree from the Institute for Digital Communications, University of Edinburgh, March, 2007. His PhD project was fully funded by Engineering and Physical Sciences Research Council (EPSRC), UK. He is currently a postdoctoral research associate in Department of Electrical Engineering, Princeton University. His research interests include signal processing and information theory and their applications in future wireless networks.

**Chao Wang** (S'07) received the B.E. degree from University of Science and Technology of China, Hefei, China, in 2003 and the MSc degree (with distinction) from The University of Edinburgh, Edinburgh, UK, in 2005. He is currently a Ph.D. candidate at The University of Edinburgh and participates in the Delivery Efficiency Core Research Programme of the Virtual Centre of Excellence in Mobile and Personal Communications. His current research projects include Multiple Input Multiple Output (MIMO) wireless systems and cooperative communications.

**John S Thompson** received his BEng and PhD degrees from the University of Edinburgh in 1992 and 1996, respectively. From July 1995 to August 1999, he worked as a postdoctoral researcher at Edinburgh, funded by the UK Engineering and Physical Sciences Research Council (EPSRC) and Nortel Networks. Since September 1999, he has been a lecturer at the School of Engineering and Electronics at the University of Edinburgh. In October 2005, he was promoted to the position of reader. His research interests currently include signal processing algorithms for wireless systems, antenna array techniques and multihop wireless communications. He has published approximately 100 papers to date including a number of invited papers, book chapters and tutorial talks, as well as co-authoring an undergraduate textbook on digital signal processing. He is currently an editor-in-chief of IEE Proceedings on Vision, Image and Signal Processing and was the technical programme co-chair for the IEEE International Conference on Communications (ICC) 2007, which was held in Glasgow.

**H. Vincent Poor** (S'72, M'77, SM'82, F'87) received the Ph.D. degree in EECS from Princeton University in 1977. From 1977 until 1990, he was on the faculty of the University of Illinois at Urbana-Champaign. Since 1990 he has been on the faculty at Princeton, where he is the Dean of Engineering and Applied Science, and the Michael Henry Strater University Professor of Electrical Engineering. Dr. Poor's research interests are in the areas of stochastic analysis, statistical signal processing and their applications in wireless networks and related fields. Among his publications in these areas are the recent books MIMO Wireless Communications (Cambridge University Press, 2007), co-authored with Ezio Biglieri, et al, and Quickest Detection (Cambridge University Press, 2009), co-authored with Olympia Hadjiliadis.

Dr. Poor is a member of the National Academy of Engineering, a Fellow of the American Academy of Arts and Sciences, and a former Guggenheim Fellow. He is also a Fellow of the Institute of Mathematical Statistics, the Optical Society of America, and other organizations. In 1990, he served as President of the IEEE Information Theory Society, and in 2004-07 as the Editor-in-Chief of these Transactions. He is the recipient of the 2005 IEEE Education Medal. Recent recognition of his work includes the 2007 IEEE Marconi Prize Paper Award, the 2007 Technical Achievement Award of the IEEE Signal Processing Society, and the 2008 Aaron D. Wyner Distinguished Service Award of the IEEE Information Theory Society.